\documentclass[%
reprint,
superscriptaddress,
nofootinbib,
amsmath,amssymb,
aps,
prb,
floatfix,
longbibliography
]{revtex4-2}

\usepackage{graphicx}
\usepackage{dcolumn}
\usepackage{bm}
\usepackage[colorlinks=true, linkcolor=blue,urlcolor=blue,citecolor=blue]{hyperref}
\usepackage[mathlines]{lineno}

\usepackage{amssymb}

\usepackage[ignoreunlbld,norefs,nocites]{refcheck}

\usepackage{ulem}

\usepackage{natbib}

\usepackage{amsmath}

\begin{document}

\preprint{APS/123-QED}

\title{Electron correlation and confinement effects in quasi-one-dimensional quantum wires at high-density}

\author{Ankush Girdhar}
\affiliation{%
Department of Physics, Dr.\ B.\ R.\ Ambedkar National Institute of Technology, Jalandhar, Punjab - 144011, India}%

\author{Vinod Ashokan}
\email{ashokanv@nitj.ac.in}
\affiliation{Department of Physics, Dr.\ B.\ R.\ Ambedkar National Institute of Technology, Jalandhar, Punjab - 144011, India}

\author{N.\ D.\ Drummond}
\affiliation{Department of Physics, Lancaster University, Lancaster LA1 4YB, United Kingdom}%

\author{Klaus Morawetz}
\affiliation{M\"unster University
of Applied Sciences, Stegerwaldstrasse 39, 48565 Steinfurt,
Germany} \affiliation{International Institute of Physics-
UFRN, Campus Universit\'ario Lagoa nova, 59078-970 Natal, Brazil }

\author{K.\ N.\ Pathak}%
\altaffiliation{Adjunct Professor at Department of Physics, Dr.\ B.\ R.\ Ambedkar National Institute of Technology, Jalandhar, Punjab -  144011, India}
\affiliation{Centre for Advanced Study in Physics, Panjab
 University, Chandigarh - 160014, India}%

\date{\today}

\begin{abstract}
 We study the ground-state properties of ferromagnetic quasi-one-dimensional quantum wires using the quantum Monte Carlo (QMC) method for various wire widths $b$ and density parameters $r_\text{s}$. The correlation energy, pair-correlation function, static structure factor, and momentum density are calculated at high density. It is observed that the peak in the static structure factor at $k=2k_\text{F}$ grows as the wire width decreases. We obtain the Tomonaga-Luttinger liquid parameter $K_\rho$ from the momentum density. It is found that $K_\rho$ increases by about $10$\% between wire widths $b=0.01$ and $b=0.5$. We also obtain ground-state properties of finite thickness wires theoretically using the first-order random phase approximation (RPA) with exchange and self-energy contributions, which is exact in the high-density limit. Analytical expressions for the static structure factor and correlation energy are derived for $b \ll r_\text{s}<1$. It is found that the correlation energy varies as $b^2$ for $b \ll r_\text{s}$ from its value for an infinitely thin wire. It is observed that the correlation energy depends significantly on the wire model used (harmonic versus cylindrical confinement). The first-order RPA expressions for the structure factor, pair-correlation function, and correlation energy are numerically evaluated for several values of $b$ and $r_\text{s} \leq 1$. These are compared with the QMC results in the range of applicability of the theory.
\end{abstract}

\maketitle


\section{Introduction}
One-dimensional (1D) homogeneous electron gases (HEGs) are known to behave as strongly correlated systems at all densities \cite{Giuliani05,Giamarchi04}. The present work studies the ground-state properties of quasi-1D electron fluids using quantum Monte Carlo (QMC) methods and an analytical theory valid in the high-density range.

An infinitely thin wire cannot be realized experimentally. As the channel width is reduced, the electrons occupy the lowest energy subband for their transverse motion, leading to a realization of a quasi-1D electron system. The transverse confinement of an electron fluid affects the electron-electron interaction potential and thus the properties of the quasi-1D electron system. The recent advancement of fabrication technology and pursuit of obtaining narrower wires has given impetus to intense experimental and theoretical research in 1D systems.
The experimental formation of 1D nanowires on surfaces makes use of the symmetry of the substrate, which can produce 1D topographic structures. In an early work by Wang \textit{et al.}\ \cite{Wang04}, a nanowire structure was obtained by depositing a monolayer of Au on a Ge ($0 0 1$) surface \cite{Dudy17}.
More recently, nanowires with well-defined long-range order and on large scale have been obtained \cite{Schafer08,Gallaghar11}. In particular, the high-density 1D HEG can be realized experimentally in zigzag carbon nanotubes formed on SrTiO$_3$ substrates with high dielectric constant \cite{Javey02, Kim04}. The semiconductor industry is expecting to achieve the building of single-digit nanometer chips. Therefore, it is relevant to study the thickness-dependent properties of 1D wires for different confinement models. In this paper, we use a harmonic confinement model for QMC simulations and we study both harmonic and cylindrical hard-wall confinement models with the theoretical approach.

We report QMC calculations of the ground-state energy, static structure factor (SSF), pair-correlation function (PCF), and momentum density (MD) at various wire widths at high density for fully spin-polarized (ferromagnetic) 1D HEGs. The Tomonaga-Luttinger (TL) parameters are key parameters describing a TL liquid \cite{Tomonaga50,Luttinger63,Haldane81,Rajesh21}. The MD data were fitted with appropriate functions around $k \sim k_\text{F}$ to obtain the TL parameter $K_\rho$. The dependence of $K_\rho$ on wire width $b$ reveals the importance of electron confinement effects in a 1D HEG\@.
In addition, we study the wire width dependence of various ground-state properties of the interacting HEG at high densities using the first-order random phase approximation (RPA) with exchange and self energy contributions \cite{Renu12,Vinod20}. We present suitable expressions for $b$-dependent SSFs, PCFs, and correlation energies for cylindrical and harmonic-potential models of transverse confinement. The high-density first-order RPA theory is found to be in good agreement with QMC results. Analytical expressions for the SSF and correlation energy have been obtained for both confinement models in the limit $b \ll r_\text{s}<1$. The $b$-dependent analytical expression for the correlation energy in the limit $b\rightarrow 0$ at high density reduces to $-\pi^2/360$ \cite{Loos13}. In this work, we take $r_\text{s}$ and the wire width $b$ to be in units of the Bohr radius $(a_\text{B})$.

The outline of this paper is as follows. In Sec.\ \ref{QMC}, we briefly describe the QMC method and report the ground-state properties such as the PCF, SSF, correlation energy, and MD for several values of $b$. The effect of the confining potential on the TL parameter $K_\rho$ is also studied. In Sec.\ \ref{Theory}, the confinement models and the dynamic density response function in first-order RPA with exchange and self energy contributions are described. Wire width dependent expressions for the SSF and correlation energy are given in this section. The SSF, PCF, and correlation energy are calculated numerically for several wire widths for $b<r_\text{s}<1$ and compared with QMC simulations. The overall conclusions are given in Sec.\ \ref{Conclusion}.

\section{Quantum Monte Carlo simulations}
\label{QMC}
The form of Hamiltonian which is used for simulating a fully spin-polarized (ferromagnetic) $N$-electron 1D HEGs is
\begin{eqnarray}
 \hat{H}=-\frac{1}{2}\sum_{i=1}^{N}\frac{\partial^2}{\partial x_i^2}+\sum_{i<j}\tilde{V}(x_{ij})+\frac{N}{2} V_{\rm Mad},
\end{eqnarray}
where $\tilde{V}(x_{ij})$ and $V_{\rm Mad}$ denote the Ewald interaction and Madelung energy, respectively. Throughout we use Hartree atomic units (a.u.), in which $\hbar=|e|=m_\text{e}=4\pi\epsilon_0=1$ a.u.

The confinement model that we have studied is a harmonically regularized Coulomb potential in which the electrons are confined to 1D by a harmonic potential of form $V_{\perp}\left(r_{\perp}\right)=r_{\perp}^{2} / 8 b^{4}$, where $b$ is a wire width parameter and $r_{\perp}$ is the distance perpendicular to the wire. Further, we follow the single-subband approximation which states that the intersubband energy must be greater than Fermi energy for electrons to occupy the lowest subband. This condition requires $r_\text{s}>\pi b/4$. Integrating over the transverse degree of freedom, one obtains the effective potential in real space as \cite{Friesen80}
\begin{eqnarray}
	\label{harmonic_potential}
	V(x)&=&\frac{\sqrt{\pi}}{2 b}{\rm e}^\frac{x^2}{ 4b^2} {\rm erfc}\left(\frac{|x|}{ 2 b}\right).
\end{eqnarray}
For a harmonic wire, the Ewald-like interaction is calculated as \cite{Saunders94,Lee11}
\begin{eqnarray}
 \tilde{V}(x_{ij})&=&\sum^{\infty}_{m=-\infty}\bigg[ \frac{\pi}{2b} e^{(x_{ij}-mL)^2/(4b)^2} {\rm erfc}\bigg(\frac{|x_{ij}-mL|}{2b}\bigg)\nonumber\\
 & &\hspace{4em} {} -\frac{1}{|x_{ij}-mL|} {\rm erf}\bigg(\frac{|x_{ij}-mL|}{2b} \bigg)\bigg]\nonumber\\
 & &{} +\frac{2}{L}\sum^{\infty}_{n=1} {\rm E}_1[(bGn)^2]\cos(Gnx_{ij}),
\end{eqnarray}
where $b$ is the wire width, $G=2\pi/L$, and ${\rm E}_1$ is the exponential integral function.
The electrostatic potential at one electron due to its interaction with all its periodic images (excluding itself) is the Madelung constant
\begin{eqnarray}
 V_{\rm Mad}=\lim_{x\rightarrow 0}\bigg[ \tilde{V}(x)-V(0)\bigg].
\end{eqnarray}
The variational and diffusion quantum Monte Carlo (VMC and DMC) techniques as implemented in the \textsc{casino} code \cite{Needs20} are used for the computation of ground-state properties of the 1D HEG at high density. A Slater-Jastrow-backflow trial wave function \cite{Drummond04,Lopez06} is used in the calculations. We computed expectation values of quantities other than the energy by combining VMC and DMC results to form extrapolated estimates \cite{Foulkes_01}. Errors in the VMC and DMC expectation values of operators that do not commute with the Hamiltonian are linear in the error in the trial wave function; however, the errors in the extrapolated estimates of the PCF and SSF are quadratic in the error in the trial wave function.  The simulation details of Ref.\ \onlinecite{Vinod18c} are followed.

\begin{figure}
 \includegraphics[width=8.5cm]{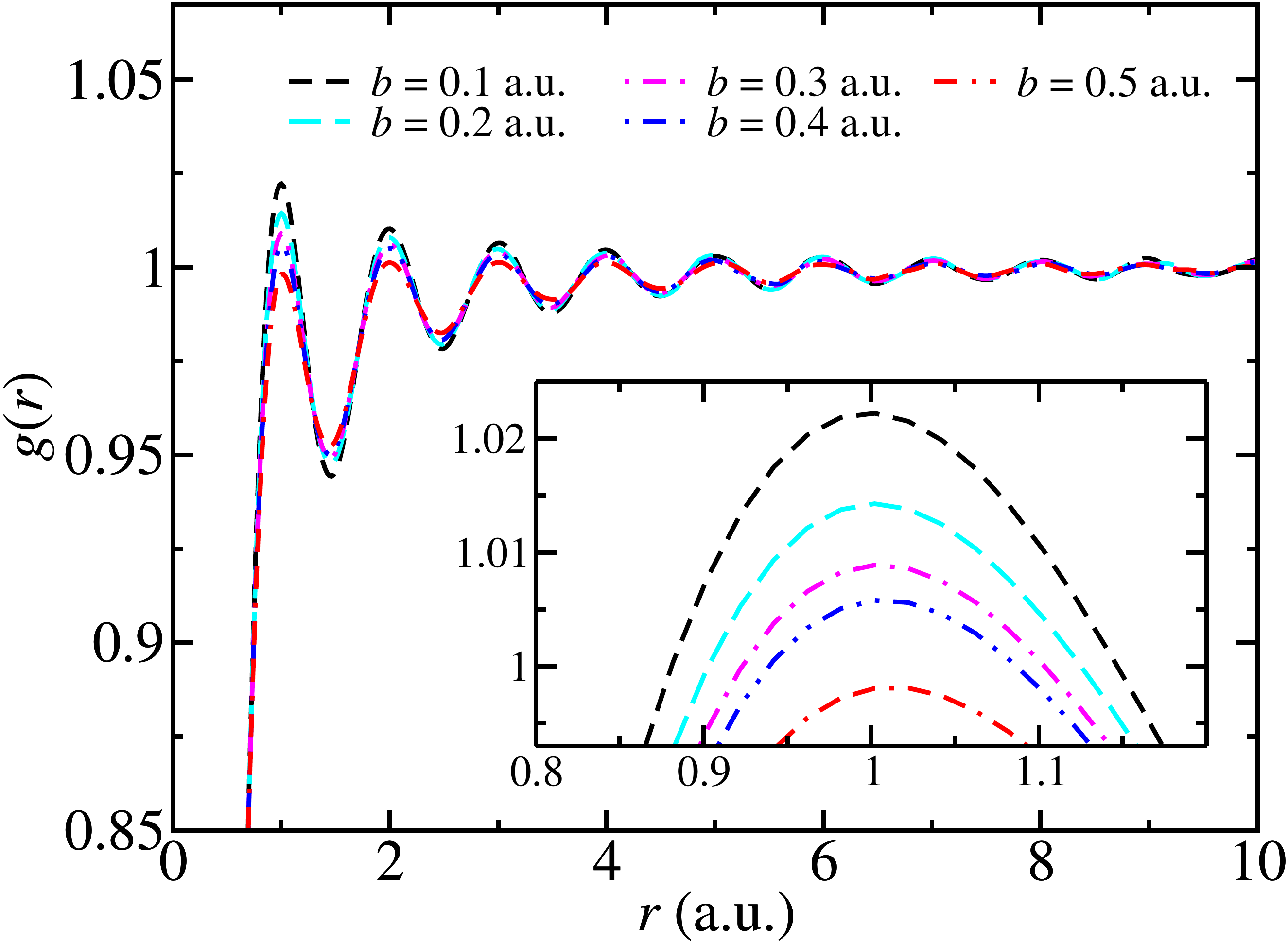}
 \caption{\label{pcf_simulation}
 	PCFs for harmonic wires with $N = 99$ and $r_\text{s}=0.5$ at various wire widths $b=0.1$ to $b=0.5$ a.u.\ (top to bottom). The inset shows a zoomed-in view of the peak at $r=1$. The data shown are extrapolated estimates ${[2g_{\rm DMC}(r)-g_{\rm VMC}(r)]}$, where $g_\text{DMC}$ and $g_\text{VMC}$ are the DMC and VMC PCFs, respectively.}
\end{figure}

\begin{figure}
 \includegraphics[width=8.5cm]{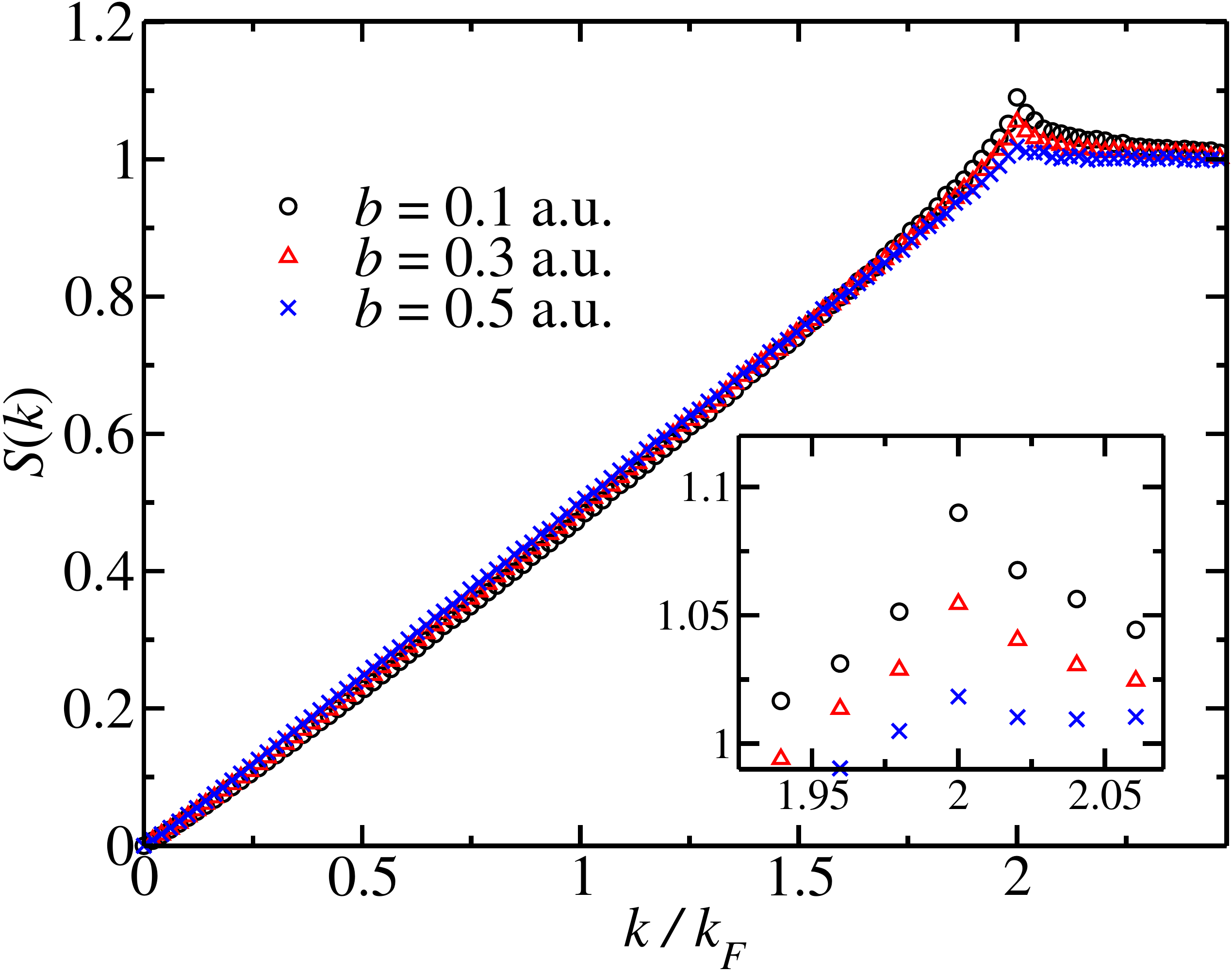}
 \caption{\label{ssf_simulation}
 	SSFs for harmonic wires with $r_\text{s}=0.5$ for various wire widths $b$. The inset shows the $b$ dependence of the $2 k_\text{F}$ peak. The data shown are for $N = 99$ and are extrapolated estimates ${[2S_{\rm DMC}(k)-S_{\rm VMC}(k)]}$, where $S_\text{DMC}$ and $S_\text{VMC}$ are the DMC and VMC SSFs, respectively.}
\end{figure}
The ground-state energy is calculated for electron numbers $N=37$, 55, 77, and 99 for harmonic wires at high densities. The thermodynamic limit for the ground-state energy per particle is obtained by extrapolating the energies per particle $E(N)$ using the fitting function $E(N)=E_{\infty}+BN^{-2}$ \cite{Lee11}, where $B$ and $E_{\infty}$ are fitting parameters. Further, the correlation energies are calculated using the DMC energies. Both are reported in Table \ref{gs_energy}.

The parallel-spin PCF is
\begin{eqnarray}
 g(r)=\frac{1}{N\rho}\bigg<\sum^{N}_{i>j}\delta(|x_{i}-x_{j}|-r)\bigg>,
\end{eqnarray}
where $\rho$ is the electron density and $x_{i}$ is the position of the $i^\text{th}$ electron. The angular brackets $\langle \cdots \rangle$ denote an average over configurations distributed as the square modulus of the wave function. The PCF of a harmonic wire at density $r_\text{s}=0.5$ for several wire widths $b$ is plotted in Fig.\ \ref{pcf_simulation}.

The SSF is defined as
\begin{eqnarray}
 S(k)=\frac{1}{N}\left[ \big< \hat{\rho}(-k)\hat{\rho}(k)\big> -\left<\hat{\rho}(-k)\right> \left<\hat{\rho}(k)\right> \right],
\end{eqnarray}
where $\hat{\rho}(k) = \sum_{i} e^{ikx_i}$. The SSF is studied to analyze the charge ordering in the system. In Fig.\ \ref{ssf_simulation}, the SSF is plotted for several wire widths. It shows that the peak height decreases with increasing wire width.

\begin{figure}
 \includegraphics[width=8.5cm]{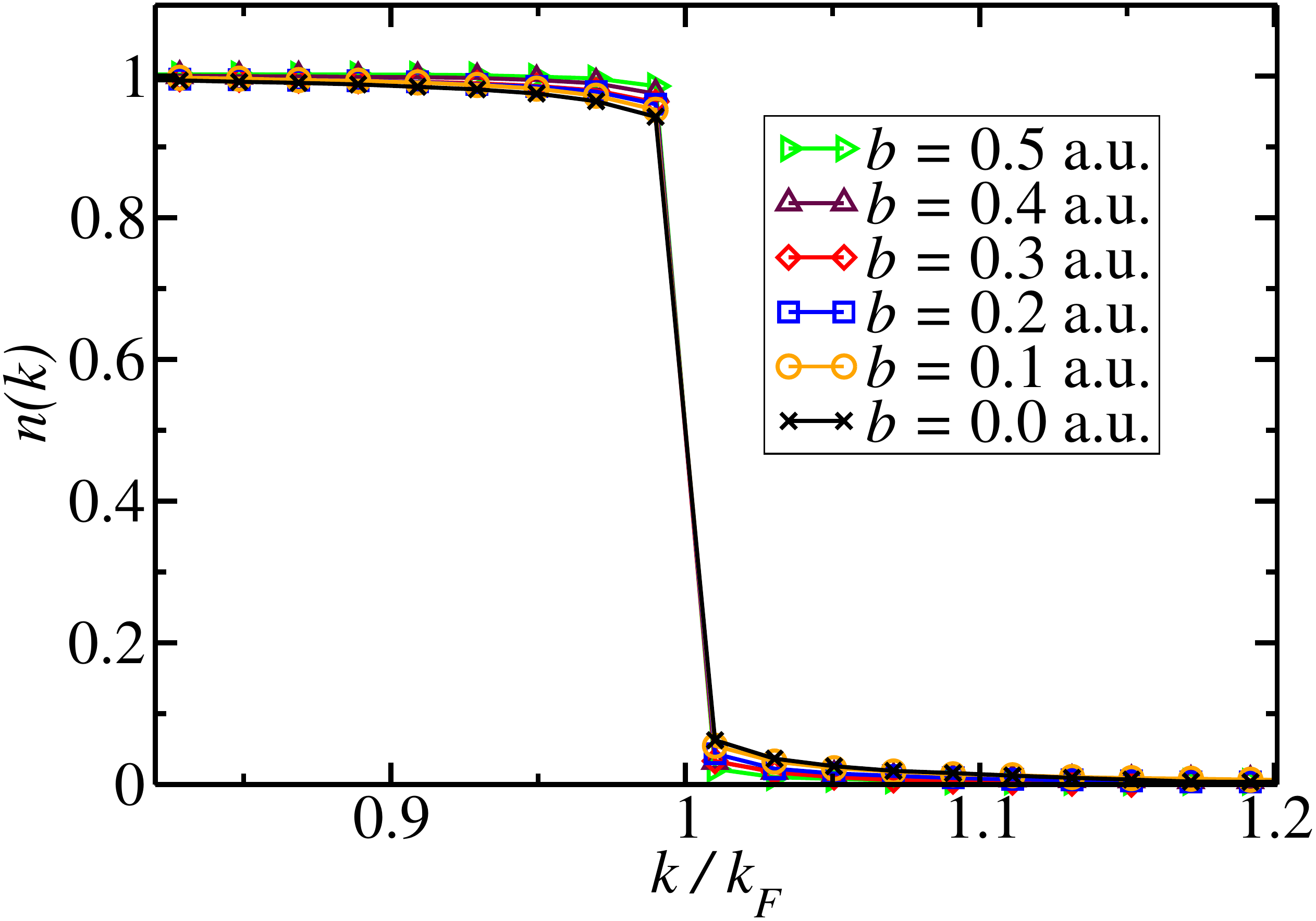}
 \caption{\label{MD_all_DMC} MDs for harmonic wires with $N = 99$ at $r_\text{s}=0.5$ for various wire widths. The data shown are extrapolated estimates ${[2n_{\rm DMC}(k)-n_{\rm VMC}(k)]}$, where $n_\text{DMC}$ and $n_\text{VMC}$ are DMC and VMC MDs, respectively. It is observed that as $b\rightarrow0$, the harmonic wire MD agrees with the infinitely thin wire MD\@. The statistical error bars are omitted for clarity as they are smaller than the symbols.}
\end{figure}

\begin{figure}
 \includegraphics[width=8.5cm]{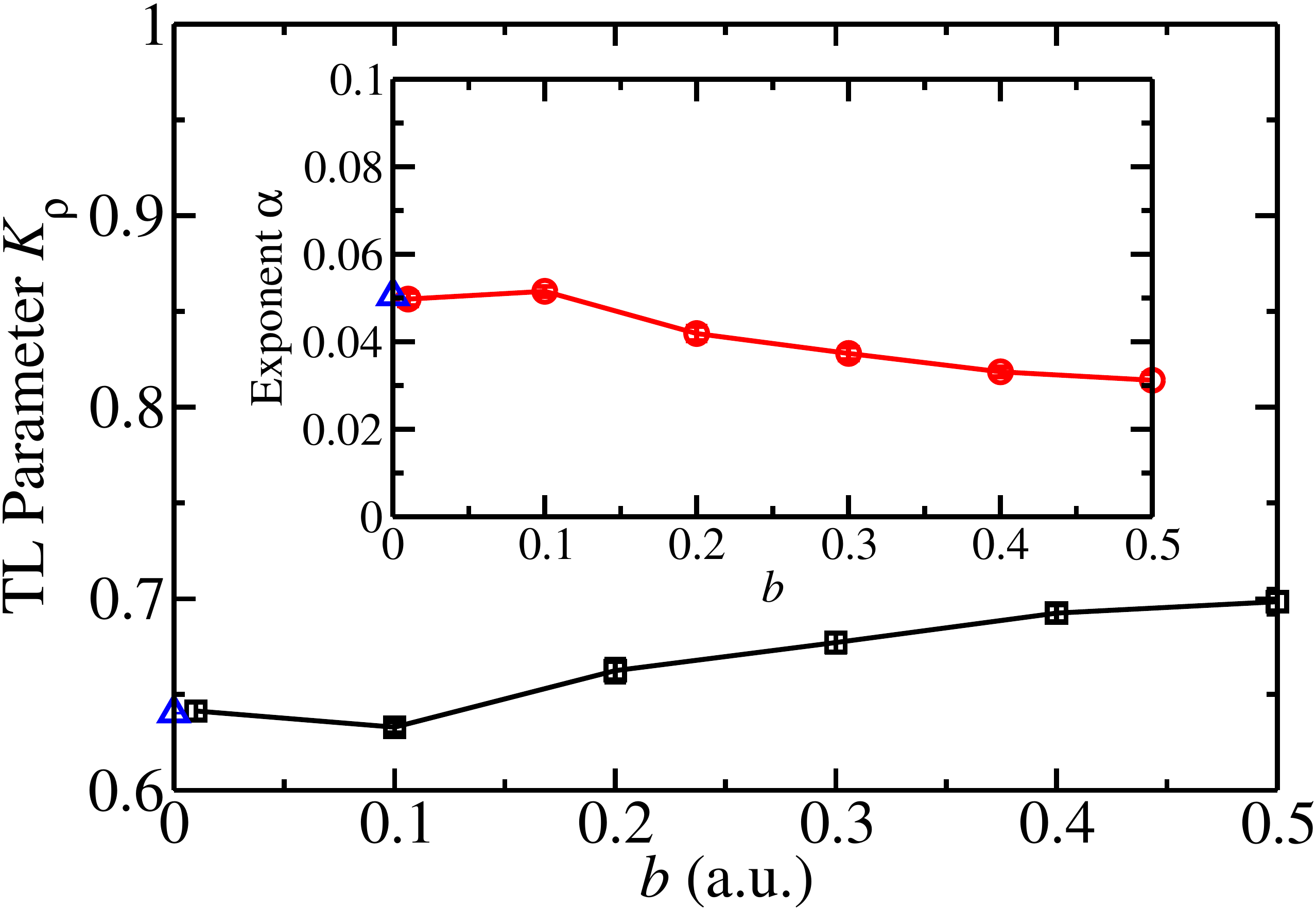}
 \caption{\label{k_rho_and_alpha} TL parameter $K_\rho$ as a function of $b$, obtained from QMC calculations. The exponent $\alpha$ is plotted against $b$ in the inset. The corresponding values for infinitely thin wires at $r_\text{s}=0.5$ are indicated on the vertical axes by the symbol `{\color{blue}$\bm{\triangle}$}'.}
\end{figure}

\begin{table}[!t]
 \begin{center}
  \caption{\label{gs_energy} DMC ground-state energies per particle extrapolated to the thermodynamic limit $E_\infty$ and correlation energies $\epsilon_\text{c}$ for fully spin-polarized (ferromagnetic) harmonic wires in the thermodynamic limit at density parameter $r_\text{s}=0.5$.}
  \begin{ruledtabular}
   \begin{tabular}{lcc}

    $b$ & $E_{\infty}$ & $\epsilon_\text{c}$\\
    \raisebox{1.5ex}[0pt]{}(a.u.) & (a.u./elec.) & (a.u./elec.) \\
    \hline
    $0.01$ & $-2.35955(6)$   & $-0.02345(6)$  \\
    $0.1$ & $-0.135195(2)$   & $-0.013913(2)$  \\
    $0.2$ & $0.434211(3)$   & $-0.007769(3) $  \\
    $0.3$ & $0.715752(3)$   & $-0.004768(3)$  \\
    $0.4$ & $0.887735(4)$    & $-0.003168(4) $ \\
    $0.5$ & $1.004636(4)$   & $-0.002239(4) $  \\
    $0.6$ & $1.089589(4)$   & $-0.001659(4) $  \\
   \end{tabular}
  \end{ruledtabular}
 \end{center}
\end{table}

The MD is computed using
\begin{eqnarray}
 n(k)=\frac{1}{2\pi}\bigg< \int \frac{\psi_\text{T}(r)}{\psi_\text{T}(x_1)}\exp[ik(x_1-r)] \, dr\bigg>,
\end{eqnarray}
where the trial wave function $\psi_\text{T}(r)$ is evaluated at $(r,x_2,\ldots,x_N)$. The angular brackets represent an average over electron configurations.

In 1D, the MD has a peculiar power-law behavior: it is continuous at $k=k_\text{F}$ although its derivative is singular at $k=k_\text{F}$. TL theory suggests that the MD takes the form \cite{Luttinger63,Mattis65}
\begin{equation}
 n(k)=n(k_\text{F})+A[{\rm sign}(k-k_\text{F})]|k-k_\text{F}|^{\alpha}
 \label{nk_MD}
\end{equation}
 near $k=k_\text{F}$, where $n(k_\text{F})$, $A$, and $\alpha$ are fitting parameters. TL theory describes the relationship between exponent $\alpha$ and TL parameter $K_\rho$ as \cite{Schulz90,Schulz93},
\begin{eqnarray}
 \alpha=\frac{1}{4}\bigg( K_{\rho}+\frac{1}{K_\rho}-2\bigg),
\end{eqnarray}
which can be rewritten as $K_\rho=1+2\alpha-2\sqrt{\alpha+\alpha^2}$. In Fig.\ \ref{MD_all_DMC}, the MD obtained using the extrapolated estimator ${2n_{\rm DMC}(k)-n_{\rm VMC}(k)}$ for the harmonic wire is plotted for $N=99$ at $r_\text{s}=0.5$ for several values of the wire width $b$. It is interesting to note that in the limit $b\rightarrow0$, the harmonic wire MD approaches the MD for an infinitely thin wire \cite{Vinod18c}. The interaction exponent $\alpha$ is calculated by fitting Eq.\ (\ref{nk_MD}) to MD data. However, we cannot use the full range of MD data for extracting $\alpha$ as Eq.\ (\ref{nk_MD}) is valid only for $k \rightarrow k_\text{F} $. So, we calculate $\alpha$ by choosing MD data in the range defined by $|k-k_\text{F}|<\epsilon k_\text{F}$, where $\epsilon > 0.075$. The exponent $\alpha$ is then calculated by extrapolating $\alpha$ to $\epsilon=0$. In this work, we report the thermodynamic value of $\alpha$ calculated by extrapolating it using $\alpha(N)=\alpha_{\infty} + B/N $ where $\alpha_{\infty}$ and $B$ are fitting parameters. Further, we calculate $K_\rho$ in thermodynamic limit. Both $\alpha$ and $K_\rho$ are plotted against $b$ in Fig.\ {\ref{k_rho_and_alpha}}. It shows that the TL parameter $K_\rho$ depends significantly on the width of the wire.

\section{Theory}
\label{Theory}

\subsection{Confinement models}
We consider a softened Coulomb potential of the form $V(x)={1}/{\sqrt{x^2+ b^2}}$, where $b$ is the transverse width parameter of the cylindrical wire. The Fourier transform of this potential is $2 K_0( b q)$ and its series expansion in $b$ is given as
\begin{eqnarray}
{V}(q)&=&-2 \left[\ln \left(\frac{ b q}{2}\right)
+\gamma \right]
\nonumber\\
&&
-\frac{ b^2q^2}{2}
\left[\ln \left(\frac{b q}{2}\right)+\gamma
-1\right]+O\left(b^3\right),
\label{v1}
\end{eqnarray}
where $K_{0}$ is the modified Bessel function of $2^\text{nd}$ kind and $\gamma$ is the Euler constant.

The second model that we have studied is a harmonic potential which is discussed in detail in Sec.\ \ref{QMC}. The Fourier transform of the potential as in Eq.\ (\ref{harmonic_potential}) is $V(q)=E_{1}( b^2 q^2)~ e^{b^2 q^2}$. Its series expansion in $b$ reads
\begin{eqnarray}
{V}(q)&=&
-2\left[  \ln ( b q)+\frac{\gamma}{2}\right]
\nonumber\\
&&
- b^2 q^2 \left[2 \ln ( b q)+\gamma -1\right]+O\left( b^3\right),
\label{v2}
\end{eqnarray}
where $E_{1}$ is the exponential integral. The series given in Eqs.\ (\ref{v1}) and (\ref{v2}) are useful only for $b<r_\text{s}$.

\subsection{Density response function}
In this section, the static properties of the 1D HEG have been obtained using the density response function and the fluctuation-dissipation theorem. The density response function is defined as \cite{Giuliani05, Renu12, Vinod20}
\begin{eqnarray}
	\label{drf_exact}
	\chi(q, \omega)=\frac{\chi_{0}(q, \omega)+\lambda \chi_{1}(q, \omega)}{1-\lambda V(q)\left[\chi_{0}(q, \omega)+\lambda \chi_{1}(q, \omega)\right]},
\end{eqnarray}
where $\lambda$ indicates the order of the potential and $\chi_{1}(q, \omega) = \chi_{1}^\text{se}(q, \omega) + \chi_{1}^\text{ex}(q, \omega)$ is the first-order correction to the polarizability with exchange and self-energy contributions.
The first-order approximation of Eq.\ (\ref{drf_exact}), valid for the high-density limit, is given as
\begin{eqnarray}
 \label{resHDE}
 \chi(q,\omega)&\approx&\chi_{0}(q,\omega)+\lambda\; v(q)\chi_{0}^2(q,\omega)\nonumber\\
 & &+\lambda\; \chi_{1}^\text{se}(q,\omega)+\lambda \;\chi_{1}^\text{ex}(q,\omega),
\end{eqnarray}
where the parameter $\lambda$ indicates the order of the expansion. The noninteracting polarizability is explicitly given as
\begin{eqnarray}
 \chi_{0}(q, \omega)=\frac{g_{s} m}{2 \pi q}
 \ln \bigg[\frac{\omega^2-(\frac{q^2}{2 m}-\frac{q
   k_{F}}{m})^2}{\omega^2-(\frac{q^2}{2 m}+ \frac{q k_{F}}{m})^2}\bigg],
\end{eqnarray}
and the self-energy and exchange contributions respectively read after simplification \cite{Vinod20}
\begin{eqnarray}
\chi_{1}^\text{se}(q,\omega)&=&2g_\text{s}\sum_{k,p}n_k n_p [v(k-p)-v(k-p+q)]\nonumber\\
&
&\times \frac{\Omega^2_{k,q}+\omega^2}{(\Omega^2_{k,q}-\omega^2)^2}
\end{eqnarray}
and
\begin{eqnarray}
 \chi_{1}^\text{ex}(q,\omega)&=&-2g_\text{s}\sum_{k,p}\bigg\{v(k\!-\!p) [n_{k-\frac{q}{2}}n_{p-\frac{q}{2}}\!-\!n_{k-\frac{q}{2}}n_{p+\frac{q}{2}}]\nonumber\\
 & &\times\frac{(\Omega_{k-\frac{q}{2},q}\;\Omega_{p-\frac{q}{2},q}+ \omega^2)}{(\Omega^2_{k-\frac{q}{2},q}-\omega^2)(\Omega^2_{p-\frac{q}{2},q}- \omega^2)}\bigg\}.
\end{eqnarray}
Here $\Omega_{k,q}=\omega_k-\omega_{k+q}$, $\Omega_{p,q}=\omega_p-\omega_{p+q}$, $g_\text{s}$ is the spin degeneracy factor, and $n_k$ denotes the Fermi-Dirac distribution function. The expressions above are directly used in our calculation.

\subsection{Structure factor}
The SSF is defined as
\begin{eqnarray}
 \label{ssf}
 S(q)=-\frac{1}{\pi\; \rho} \int_{0}^{\infty} d\omega\; \chi''(q,\omega),
\end{eqnarray}
where $\chi''(q,\omega)$ is the imaginary part of the density
response function. The integral in Eq.\ (\ref{ssf}) can be rewritten using the
contour integration method \cite{Giuliani05} as
\begin{eqnarray}
 \label{ssfRPA}
 S(q)=-\frac{1}{\pi\; \rho} \int_{0}^{\infty} d\omega\; \chi(q,i\,
 \omega).
\end{eqnarray}
Substituting Eq.\ (\ref{resHDE}) into Eq.\ (\ref{ssfRPA}), the total SSF can be written as
\begin{eqnarray}
 \label{ssftotal}
 S(q)=S_0(q)+S^\text{d}_1(q)+S^\text{se}_1(q)+S^\text{ex}_1(q).
\end{eqnarray}
The noninteracting SSF is given as $S_0(q)=x$ for $x<1$ and $S_0(q)=1$ for $x>1$ with $x=q/2k_\text{F}$.
The first-order SSF is defined as $S_1(x)=S_1^\text{d}(x)+S_1^\text{se}(x)+S_1^\text{ex}(x)$.
 The $\omega$ integration of the self-energy term turns out to be zero so that there is no contribution of $S_{1}^\text{se}(x)$ to the SSF\@. The sum of both corrections, $S_1^\text{d}(x)$ and $S_1^\text{ex}(x)$ for small $b$, and next term in series expansion are denoted $S^\text{Hr.}_1(x)$ and $S^\text{Hr.}_1(x, b)$, respectively, for harmonic wires and similarly as $S^\text{Cy.}_1(x)$ and  $S^\text{Cy.}_1(x, b)$ for cylindrical wires.

\begin{figure}
 \includegraphics[width=8.5cm]{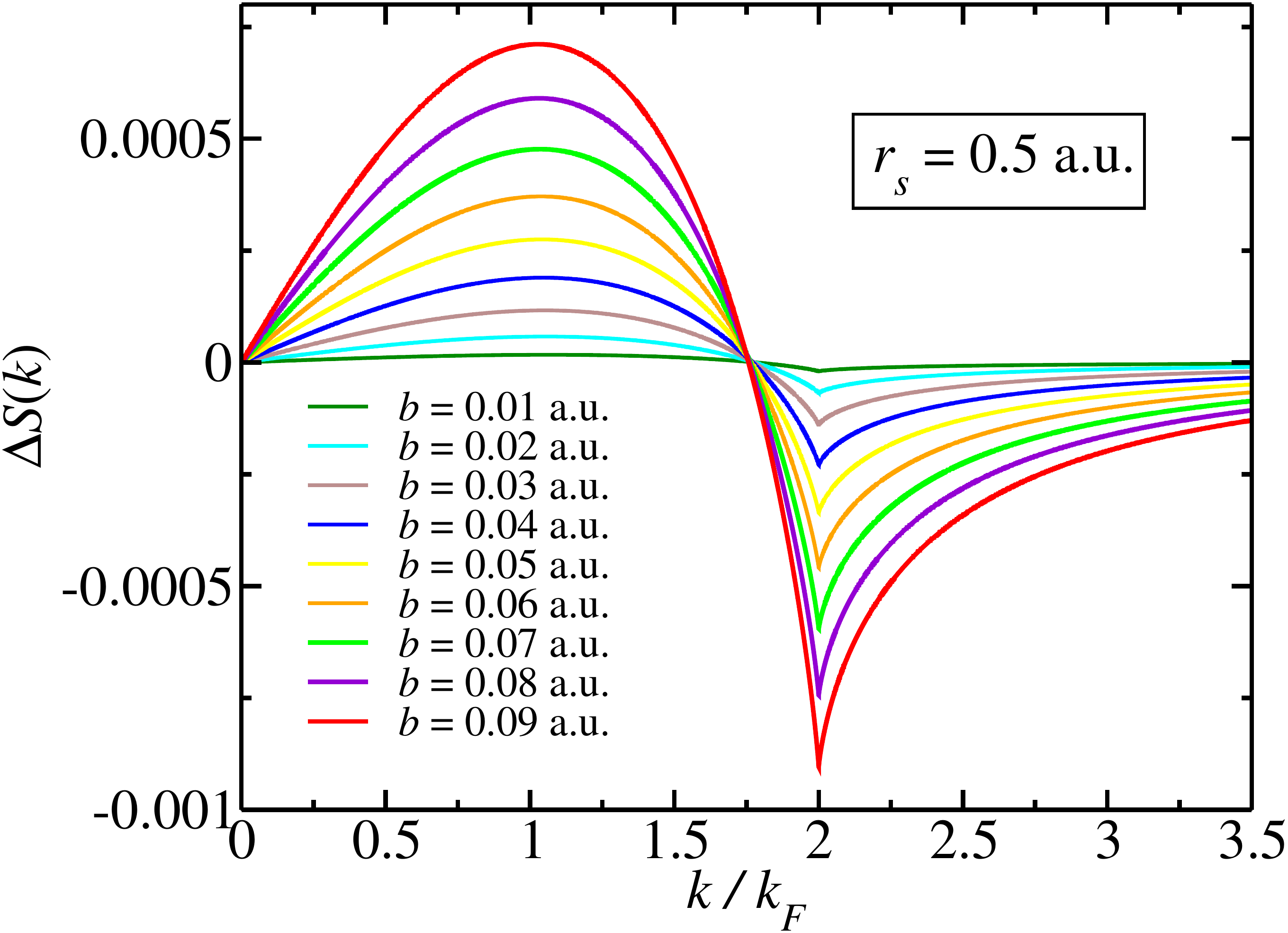}
 \caption{\label{ssf_2nd_difference} SSF difference  $\Delta S(k) = S^\text{Hr.}_1(x, b) - S^\text{Hr.}_1(x)$ as a function of $k/k_\text{F}$ for several harmonic wire widths $b=0.01$--0.09 for $r_\text{s}=0.5$, evaluated using the first-order RPA\@. The greatest value of $|\Delta S(k)|$ is for $b=0.09$ in this plot.}
\end{figure}

The exchange contribution to the SSF \cite{Vinod20} for $x<1$ is
\begin{align}
 S^\text{ex}_1(q)=\frac{g_\text{s}^2r_\text{s}}{\pi^2 x}
 &\bigg[\left ((1+x)\int\limits_{1}^{1+x}-(1-x)\int\limits_{1-x}^{1}\right )
 \frac{d \bar x}{ \bar x} {v}(\bar x)\nonumber\\
 +&\left (\int\limits_{1-x}^{1} - \int\limits_{1}^{1+x~}\right )
 d {\bar x} v(\bar x)\bigg]
 \label{Sva}
\end{align}
and similarly for $x>1$ it is
\begin{align}
 S^\text{ex}_1(q)=\frac{g_\text{s}^2r_\text{s}}{\pi^2 x}
 &\bigg[\left ( (1+x)\int\limits_{x}^{1+x}-
 (x-1)\int\limits_{x-1}^{x}\right ) \frac{d \bar x}{ \bar x} {v}(\bar x)\nonumber\\
 +&\left (\int\limits_{~x-1}^{x} - \int\limits_{x}^{1+x~}\right )
 d {\bar x} \, v(\bar x)\bigg].
 \label{Svb}
\end{align}
The explicit integrals in Eqs.\ (\ref{Sva}) and (\ref{Svb}) can be solved analytically for a given potential to calculate the exchange term. The direct term contribution \cite{Vinod18a,KM18} is obtained in the small-$b$ limit for $x<1$ as
\begin{align}
 S_1^\text{d}(x)&=-\frac{g_\text{s}^2 r_\text{s}} {\pi ^2 x} \bigg[ \bigg((1-x) \ln (1-x)\nonumber\\
 &+(x+1) \ln (x+1)\bigg)v(x)\bigg]
\end{align}
and for $x>1$ as
\begin{align}
 S_1^\text{d}(x)=&-\frac{g_\text{s}^2 r_\text{s}}{\pi ^2 x}\bigg[\bigg( (x-1) \ln (x-1)-2 x \ln (x)\nonumber\\
 &+(x+1) \ln (x+1)\bigg)v(x) \bigg].
\end{align}

 On including the next term of the series expansion $O\left( b^2\right)$ of the harmonic potential [Eq.\ (\ref{v2})], the analytical expression for the sum of $S^\text{ex}_1(q)$ and $S_1^\text{d}(x)$ is derived for $x<1$ as
\begin{align}
 S^\text{Hr.}_1(x, b)=&S^\text{Hr.}_1(x)+\frac{g_\text{s}^2 r_\text{s} b^2}{3\pi ^2 x} \bigg\{
 3 x^2(2 \ln (b x)+\gamma -1)\nonumber\\
 \times& [(x-1) \ln |x-1|+(x+1)\ln (x+1))] \nonumber\\
 -&6 x^2 \ln (b)+  x^2(-3 \gamma+8 ) \nonumber\\
 -&|(x-1)^3| \ln |x-1|-(x+1)^3\ln (x+1)\bigg\}
\end{align}
and for $x>1$,
\begin{align}
 S^\text{Hr.}_1(x, b)=&S^\text{Hr.}_1(x)+\frac{g_\text{s}^2 r_\text{s} b^2}{3\pi ^2 x} \bigg\{
 3 x^2(2 \ln (b x)+\gamma -1)\nonumber\\
 \times& [(x-1) \ln (x-1)+(x+1) \ln
 (x+1) \nonumber\\
 -&2 x \ln (x)]-18 x^3 \ln b+x \big(2 x^2 \ln (x)-3 \gamma
 +8\big) \nonumber\\
 -&(x-1)^3 \ln (x-1)-(x+1)^3 \ln (x+1)\bigg\}.
\end{align}
\begin{figure}
 \includegraphics[width=8.5cm]{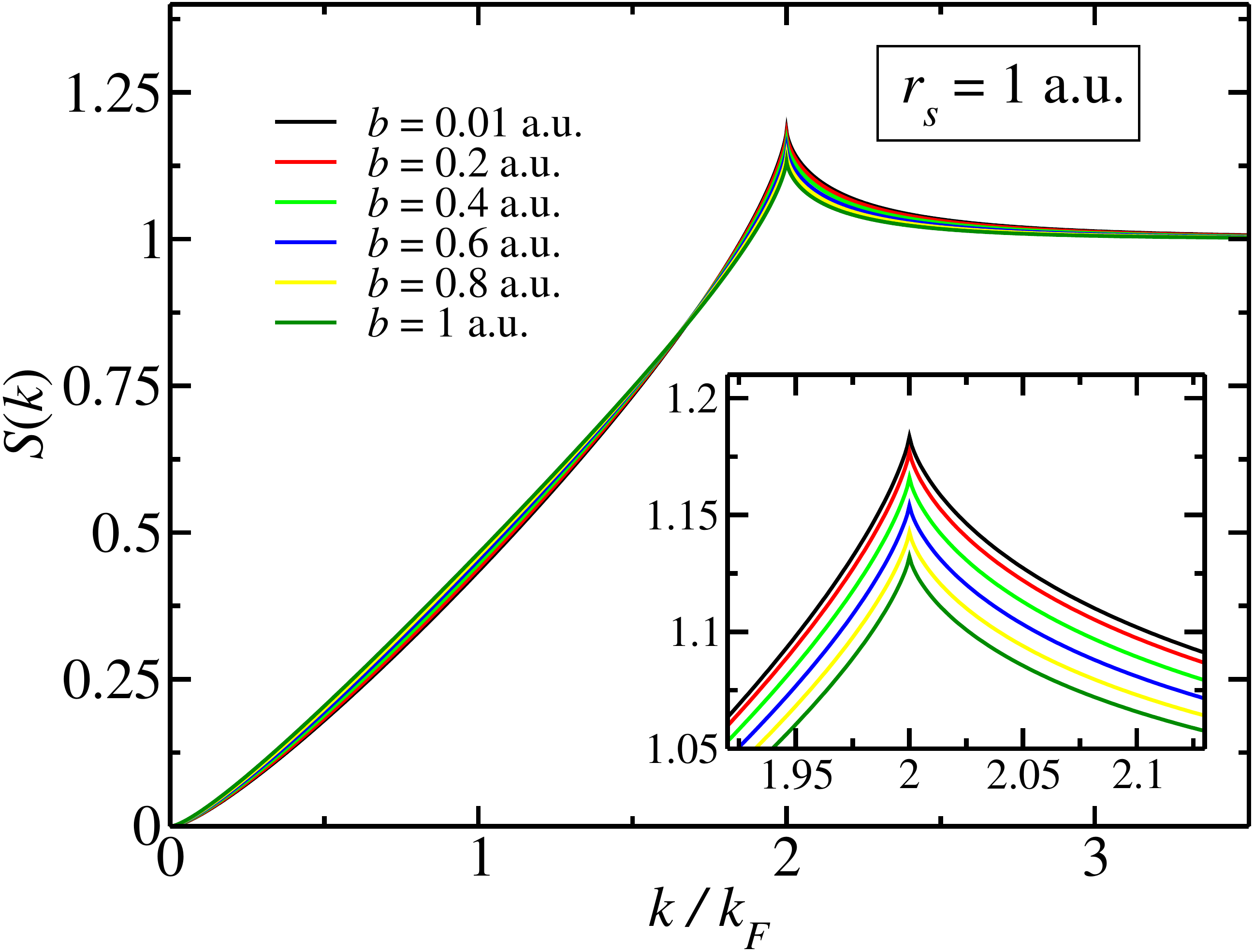}
 \caption{\label{ssf_gen_har} First-order RPA SSF as a function of $k/k_\text{F}$ for various harmonic wire widths $b=0.01$--1 (top to bottom) at $r_\text{s}=1$. The inset shows a zoomed-in view of the $2 k_\text{F}$ peak.}
\end{figure}
\begin{figure}
 \includegraphics[width=8.5cm]{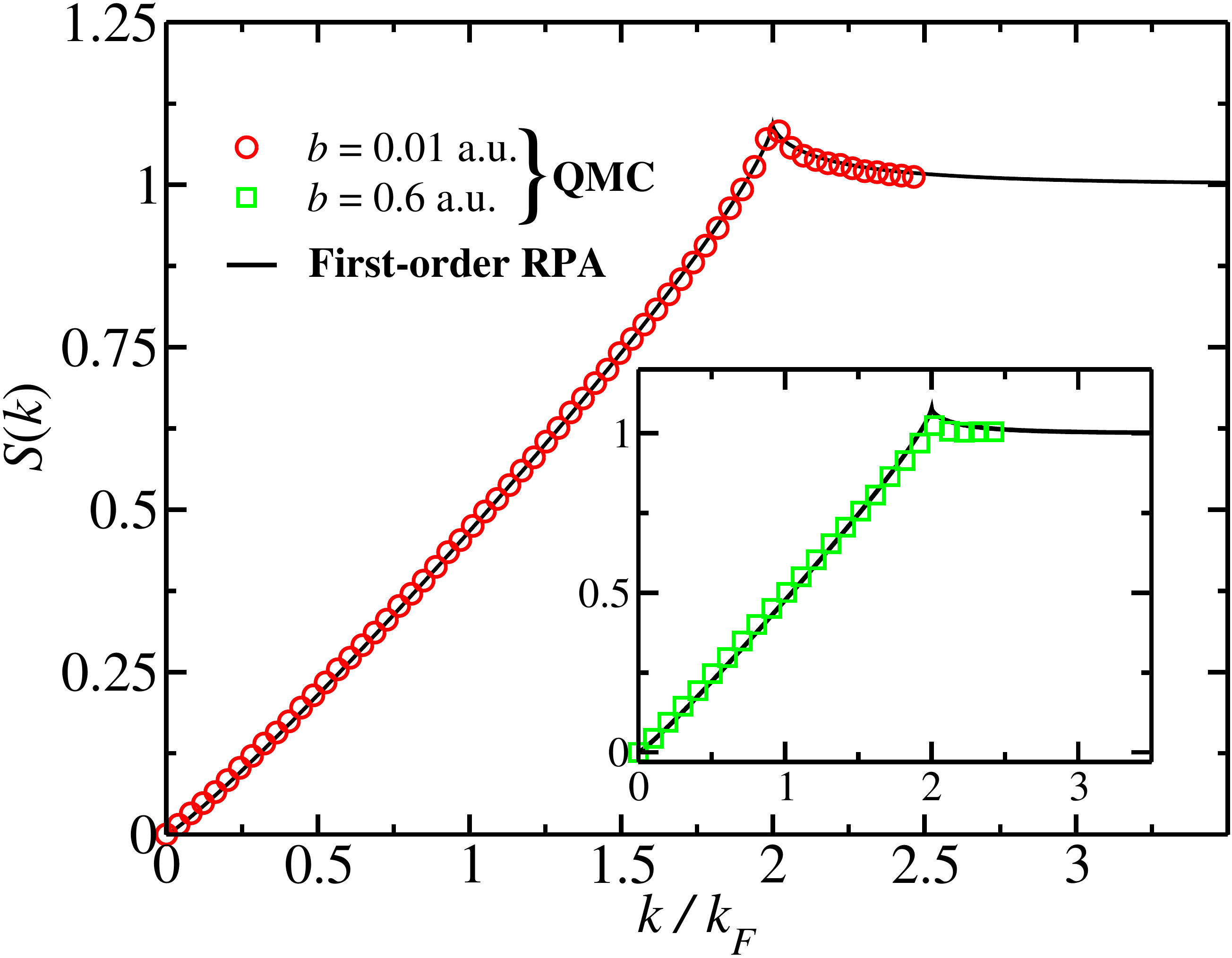}
 \caption{\label{ssf_sim_theory}
 	First-order RPA for harmonic wires (solid line) compared with extrapolated estimates ${[2S_{\rm DMC}(k)-S_{\rm VMC}(k)]}$ of SSFs for $N=99$ at $r_\text{s}=0.5$. The main plot shows the SSF for $b=0.01$, and the inset is for $b=0.6$.}
\end{figure}
We plot the difference $\Delta S(k) = S^\text{Hr.}_1(x, b) - S^\text{Hr.}_1(x)$ in Fig.\ \ref{ssf_2nd_difference} at $r_\text{s}=0.5$ for a harmonic wire. It shows that in the limit $b\rightarrow 0$, the $b$-dependent correction reduces to zero.

The expression for $S^\text{Cy.}_1(x, b)$ for the next term of series expansion $O\!\left( b^2\right)$ of the cylindrical potential (\ref{v1}) is given for $x<1$ as,
\begin{align}
 S^\text{Cy.}_1(x, b)=&S^\text{Cy.}_1(x)+ \frac{g_\text{s}^{2} r_\text{s} b^2}{12 \pi^{2} x}
 \bigg\{6 x^2 [|x-1| \ln|x-1|\nonumber\\
 -&(x+1) \ln(x+1)] \big[\ln\left(\frac{bx}{2}\right)+\gamma -1\big]\nonumber\\
 +&6 x^2 \ln\left(\frac{2}{b}\right)+(11-6 \gamma )
 x^2\nonumber\\
 -&|(x-1)^3| \ln|x-1|-(x+1)^3 \ln(x+1)
 \bigg\}
\end{align}
and for $x>1$,
\begin{align}
 S^\text{Cy.}_1(x,b)=&S^\text{Cy.}_1(x)+ \frac{g_\text{s}^{2} r_\text{s} b^2}{12\pi^{2} x}
 \bigg\{
 6 x^2 [(x-1) \ln (x-1)\nonumber\\
 -&2 x \ln (x)+(x+1) \ln (x+1)] \big[\ln \left(\frac{b x}{2}\right)+\gamma \nonumber\\
 -&1\big] + 6 x \ln \left(\frac{2}{b}\right)
 +x \left[2 x^2 \ln (x)-6 \gamma +11\right]\nonumber\\
 -&(x-1)^3 \ln(x-1)-(x+1)^3 \ln(x+1)
 \bigg\},
\end{align}
where the expressions for $S^\text{Hr.}_1(x)$ and $S^\text{Cy.}_1(x)$ for $x<1$ and $x>1$ are presented in Appendix \ref{App:ssf_b_indep}.

The SSF of Eq.\ (\ref{ssftotal}) for a harmonic wire is calculated numerically and plotted in Fig.\ \ref{ssf_gen_har} for various wire widths. The SSF shows a peak at $2k_\text{F}$, which is illustrated in greater detail in the inset of the figure. The peak height is seen to increase in the limit $b \rightarrow 0$. Further, in Fig.\ \ref{ssf_sim_theory}, we compare the SSF with our QMC data. The wire width dependence of the SSF shows excellent agreement with our first-order RPA results.

\subsection{Pair-correlation function}
\begin{figure}
 \includegraphics[width=8.5cm]{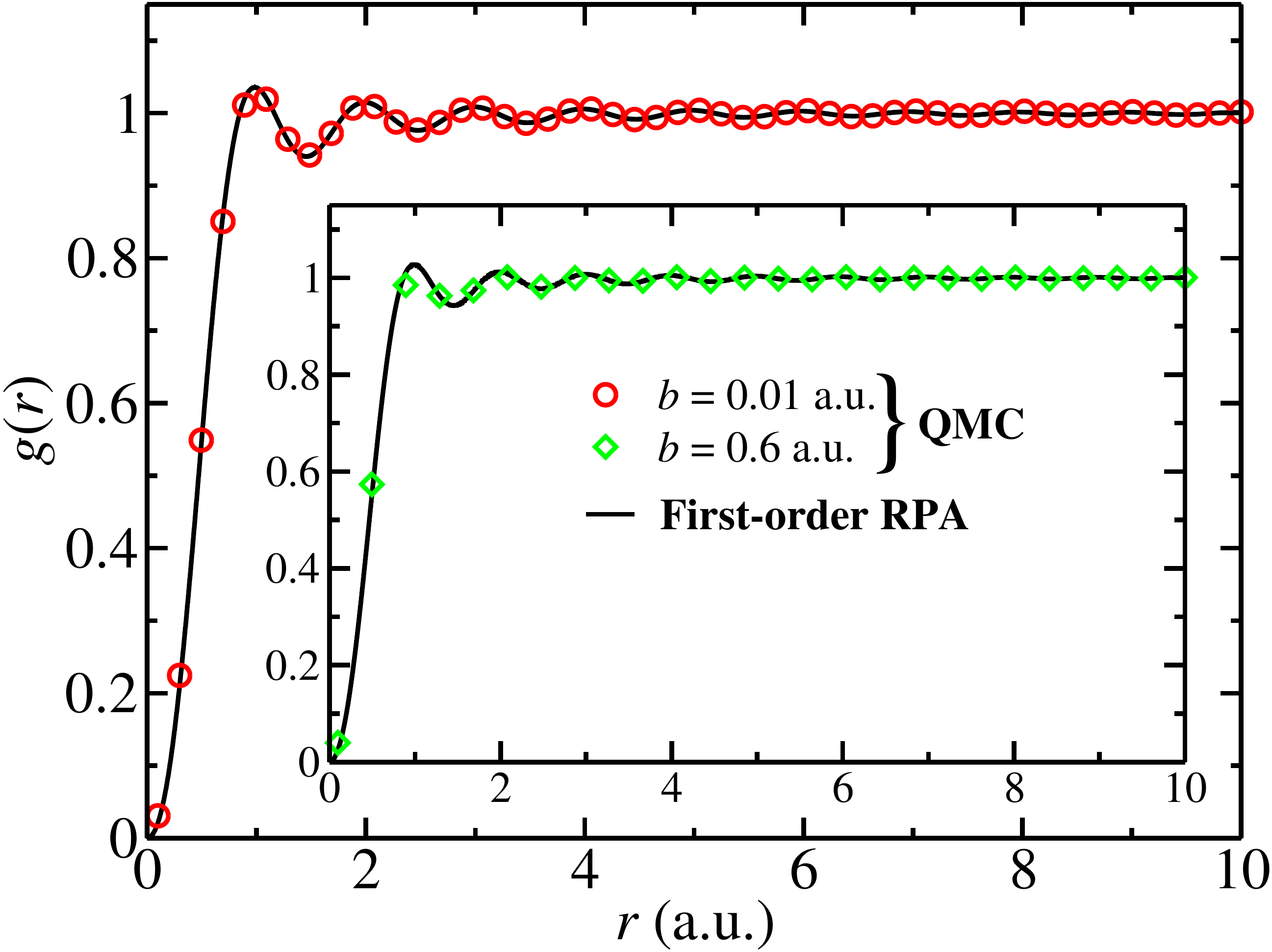}
 \caption{\label{pcf_sim_theory} Theoretical PCF compared with extrapolated estimates of PCFs ${[2g_{\rm DMC}(r)-g_{\rm VMC}(r)]}$ for harmonic wires with $N=99$ at $r_\text{s}=0.5$. The main plot is for $b=0.01$ and the inset is for $b=0.6$.}
\end{figure}

The PCF $g(r)$ is obtained from the SSF $S(q)$ as
\begin{eqnarray}
 \label{pcf_formula}
 g(r)=1-\frac{1}{2\pi \rho}\int^{\infty}_{-\infty} dq\; e^{iqr}[1-S(q)].
\end{eqnarray}
 In Fig.\ \ref{pcf_sim_theory}, the PCF evaluated using Eq.\ (\ref{pcf_formula}) is plotted for a finite-width harmonic wire. The confinement effect of correlations in the PCF is compared with the QMC simulation results and found to be in very good agreement.

\subsection{Ground-state energy}
The ground-state energy in terms of the density-density response function can be written using the fluctuation-dissipation theorem \cite{Vinod18a} as
\begin{eqnarray}
\label{gsE}
E_\text{g}&=&E_0+\frac{\rho}{2}\sum_{q\neq 0}V(q)\nonumber\\
& \times& \bigg( -\frac{1}{\rho\pi}\int^1_0 d\lambda \int^{\infty}_0 \chi(q,\iota \omega;\lambda)\; d\omega-1\bigg),
\end{eqnarray}
where $\rho=(k_\text{F}~g_\text{s})/\pi$ is the linear electron number density and $k_\text{F}$ is the Fermi wave vector. Using Eq.\ (\ref{resHDE}) in Eq.\ (\ref{gsE}), the ground-state energy can be written as a sum of the kinetic energy of the noninteracting HEG $E_0$, the exchange energy $\epsilon_\text{x}$, and the correlation energy $\epsilon_\text{c}$ as
\begin{eqnarray}
 \label{gse_all}
 E_\text{g}=E_0+E_\text{x}+E_\text{c},
\end{eqnarray}
where the kinetic energy is given by $E_0={\pi ^2}/{24 r_\text{s}^2}$ for the fully spin-polarized HEG\@. The exchange energy contribution is given by
\begin{align}
E_\text{x}&=\frac{\rho}{2}\sum_{q\neq 0}V(q)\bigg( -\frac{1}{\rho\pi}\int^1_0 d\lambda \int^\infty_0 \chi_0(q,\iota\omega)d\omega-1\bigg)\nonumber\\
&=\frac{\rho}{2}\sum_{q\neq 0}V(q) [S_0(q)-1)],
\end{align}
and the correlation energy is
\begin{align}
E_\text{c}&=\frac{\rho}{2}\sum_{q\neq 0}V(q)\bigg( -\frac{1}{\rho\pi}\int^1_0 d\lambda \int^\infty_0 \bigg\{ \lambda\; V(q) \chi_0^2(q,\iota\omega)\nonumber\\
& +\lambda\; \chi_1^\text{se}(q,\iota\omega)+\lambda\; \chi_1^\text{ex}(q,\iota\omega)\bigg\}d\omega\bigg)\nonumber\\
&=\frac{\rho}{4}\sum_{q\neq 0}V(q) [S^\text{d}_1(q)+S_1^\text{se}(q)+S_1^\text{ex}(q)]\nonumber\\
\epsilon_\text{c}&=\frac{1}{4\pi}\int_{0}^{\infty} V(q) [S^\text{d}_1(q)+S_1^\text{se}(q)+S_1^\text{ex}(q)] dq.
\label{eq:corr_e_analytic}
\end{align}

\begin{figure}
 \includegraphics[width=8.5cm]{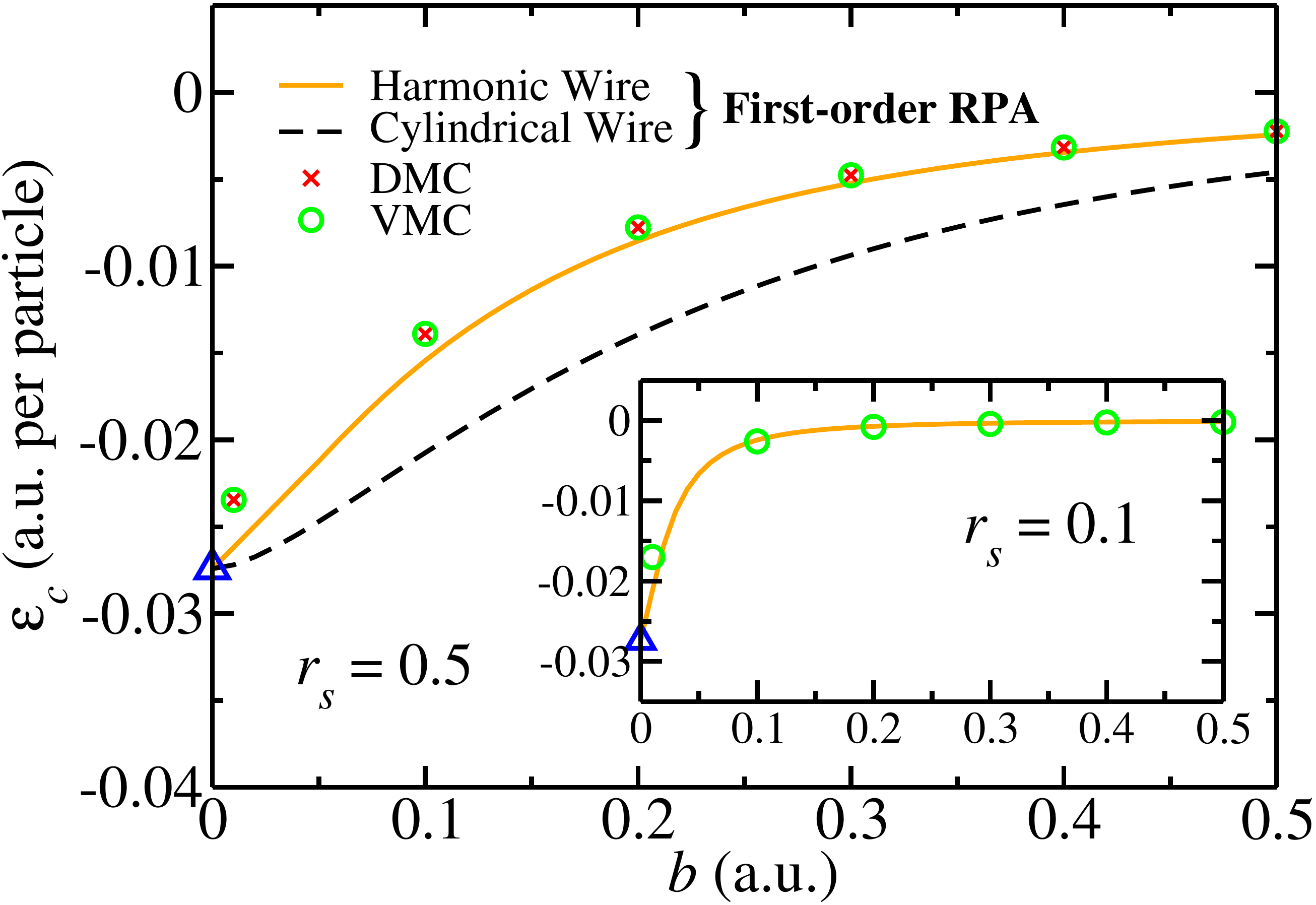}
 \caption{\label{ce_all} Correlation energy per particle for a harmonic wire and a cylindrical wire calculated using first-order RPA (shown as solid and dashed lines) valid in the high-density limit, compared with VMC and DMC data obtained for a harmonic wire. The main plot shows the comparison for $r_\text{s}=0.5$ and the inset is for $r_\text{s}=0.1$. The value of the first-order RPA correlation energy for an infinitely thin wire is represented by `{\color{blue}$\bm{\triangle}$}'. The disagreement of the QMC data with first-order RPA at smaller values of $b$ is due to the approximation used in Eq.\ (\ref{resHDE}), which restricts its applicability to the high-density limit.}
\end{figure}

\begin{figure}
 \includegraphics[width=8.5cm]{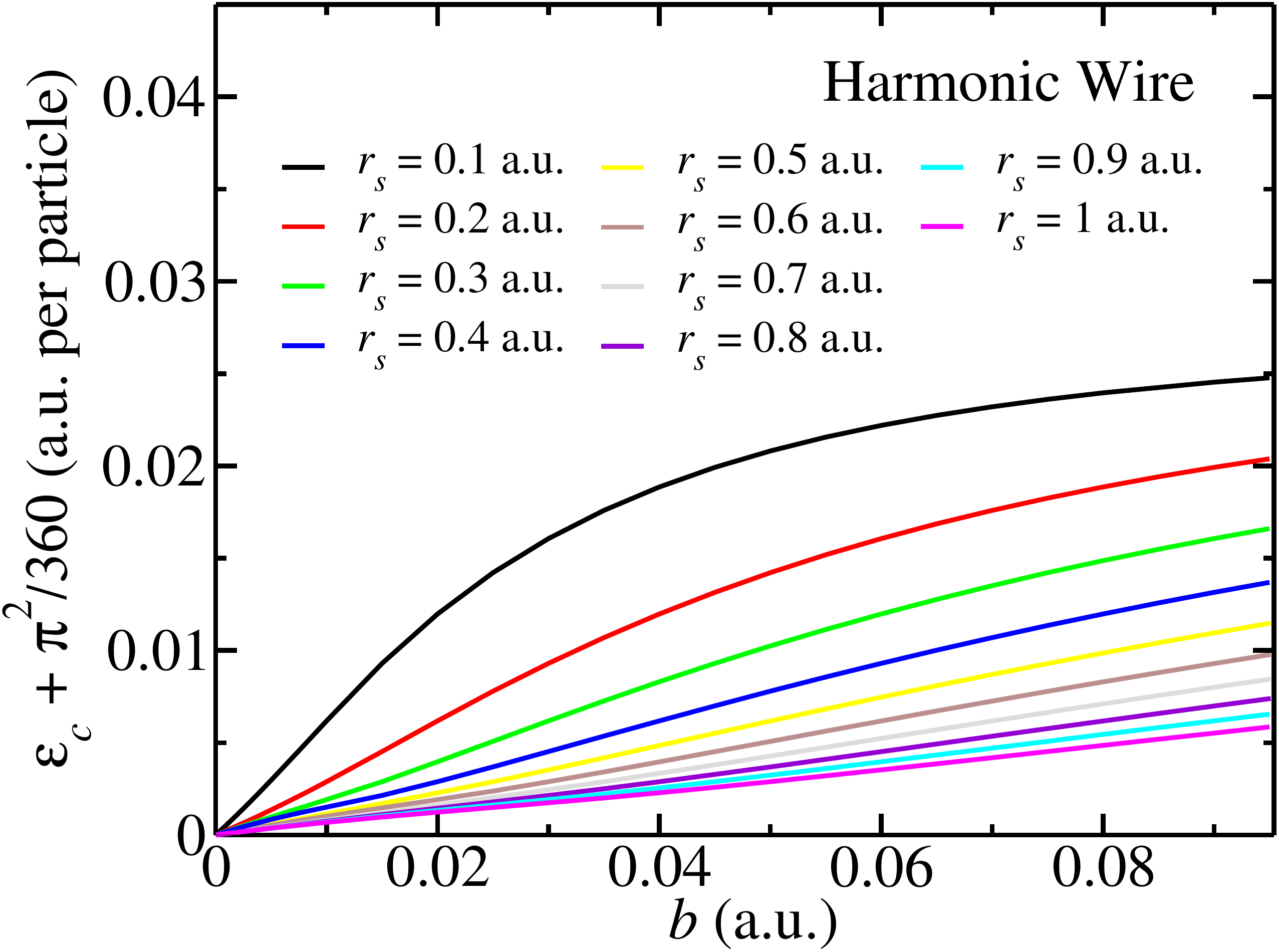}
 \vskip 0.2in
 \includegraphics[width=8.5cm]{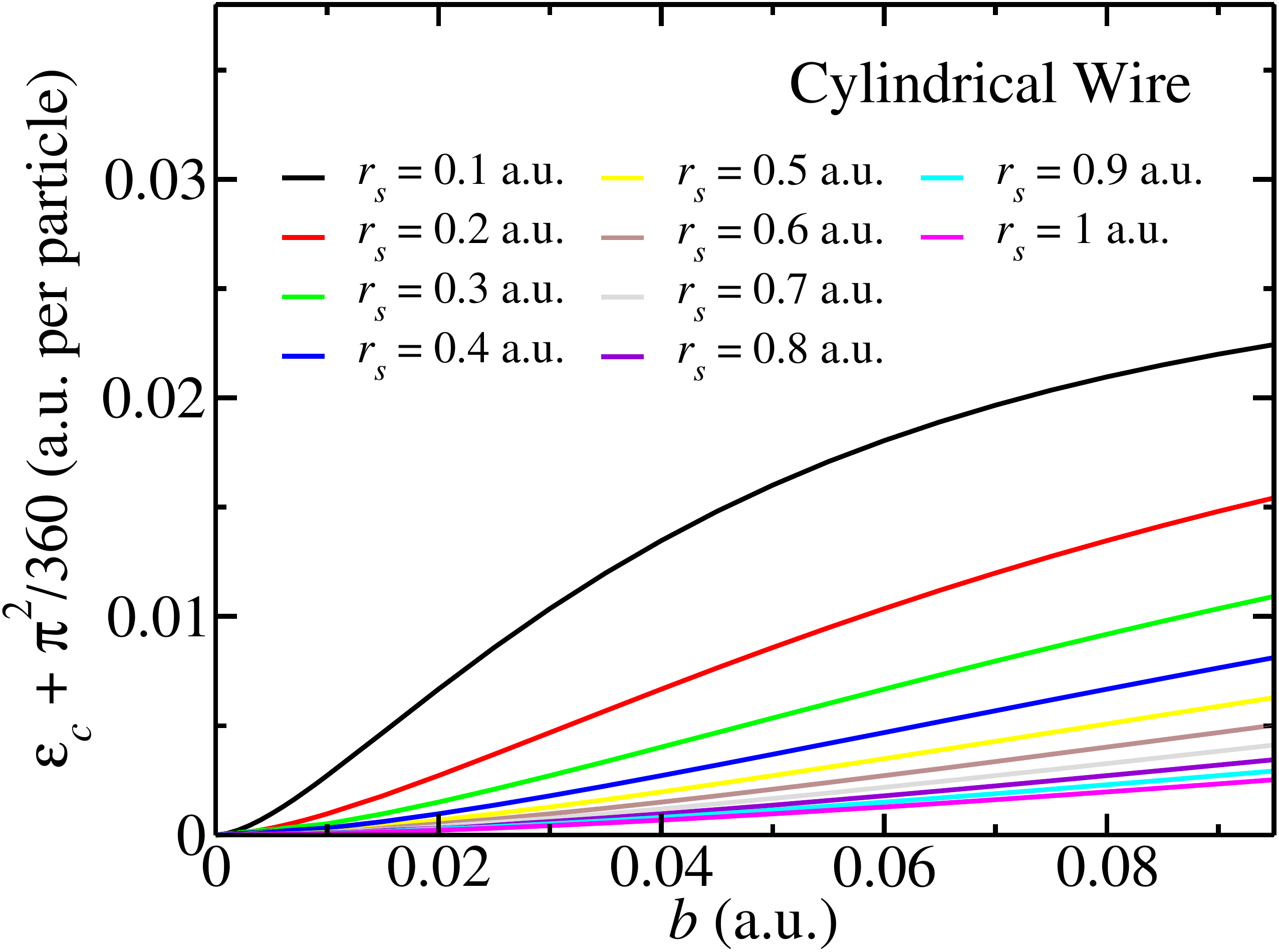}
 \caption{\label{corr_gen_both} Change in the first-order RPA correlation energy per particle from its value for the infinitely thin wire as a function of $b$ for a harmonic wire (upper) and cylindrical wire (lower) for various $r_\text{s}$ values ($0.1$--$1$) from top to bottom.}
\end{figure}

 We calculate the correlation energy as in Eq.\ (\ref{eq:corr_e_analytic}) numerically within the range of applicability of our theory for both harmonic and cylindrical wires. The results are shown in Fig.\ \ref{ce_all}. In this figure, we also present result for $r_\text{s}=0.1$ in the inset. It is seen that the difference between QMC and theoretically calculated values decreases as $r_\text{s}$ is made smaller. The difference in the two values for $r_\text{s}=0.5$ at $b=0.01$ is about 10\%. The QMC data obtained for a harmonic potential are also shown as symbols. There is a significant difference between the correlation energies of harmonic and cylindrical wires as $b$ increases. The difference of the correlation energy from its value for the infinitely thin wire is plotted against the wire width in Fig.\ \ref{corr_gen_both}.

\begin{figure}
	\includegraphics[width=8.5cm]{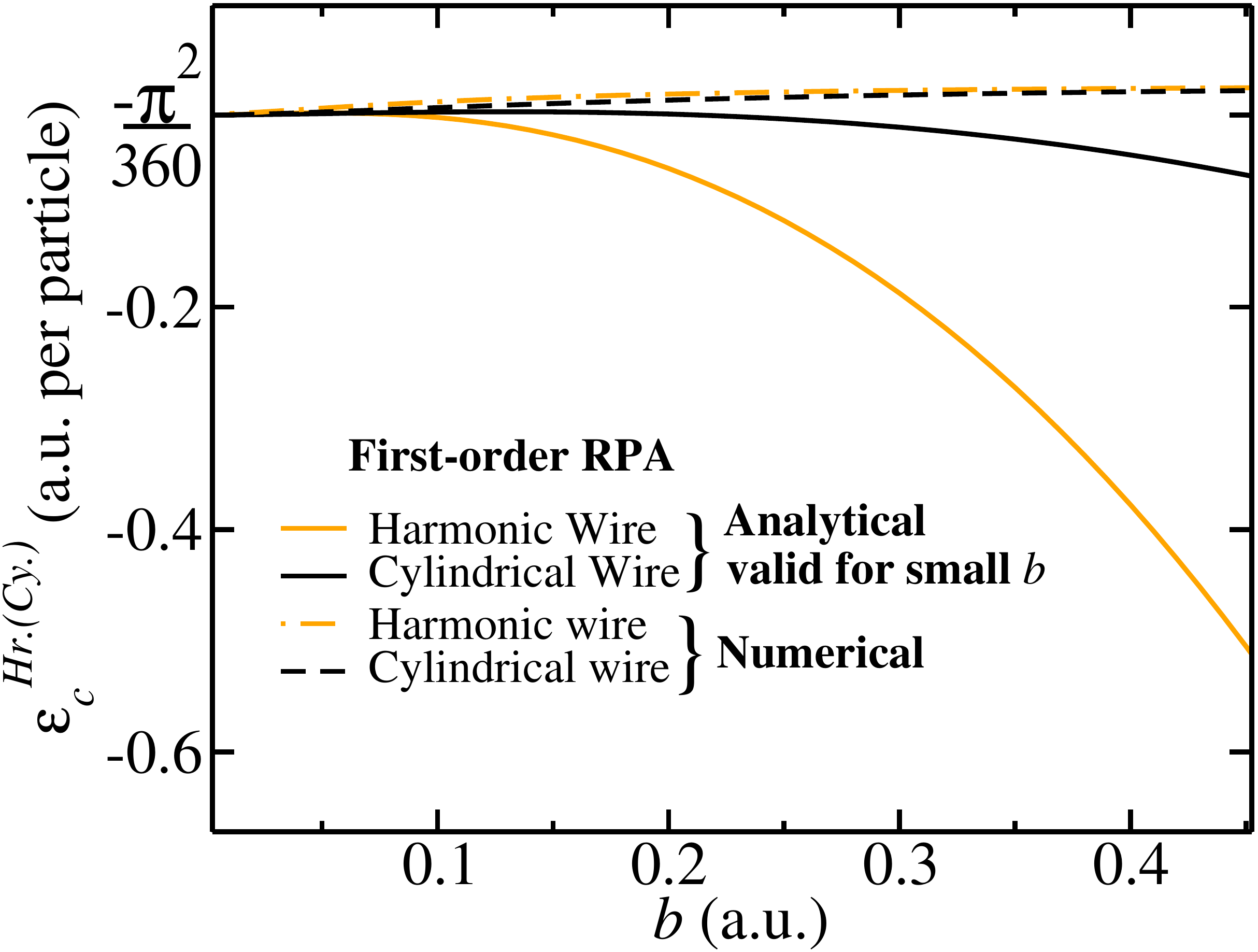}
	\caption{\label{ce_2nd_order} Analytical expressions for the wire-width dependent correlation energy [Eqs.\ (\ref{exactH}) and (\ref{exactC})], which are valid only for small $b$, plotted as a function of wire width $b$ for $r_\text{s}=0.5$. The solid curves are for analytical results and dashed curves are result of numerical calculations of correlation energy as in Eq.\ (\ref{eq:corr_e_analytic}). The plot shows the applicability of the analytical results in the range $b<0.1$.}
\end{figure}

An analytical expression for the wire-width-dependent correlation energy is derived using the next term of the series expansion $O\left( b^2\right)$ of the potential [Eq.\ (\ref{v2})] for a harmonic wire and the SSF as in Eqs.\ (\ref{SH1a}) and (\ref{SH1b}) for $x<1$ and $x>1$, respectively. The correlation energy per particle reads
\begin{eqnarray}
 \epsilon_\text{c}^\text{Hr.}(b, r_\text{s})
 &=&\frac{1}{4r_\text{s}}\bigg\{\Lambda_{1}+\Lambda_{2} \bigg\},
\end{eqnarray}
where $\Lambda_{1}$, the contribution for $x<1$, is
\begin{align}
		\label{lamda1H}
		 \Lambda_{1}=&\int^1_0 v(x) [S^\text{Hr.}_1(x)]_{x<1}\;dx
		 \nonumber\\
		=&\frac{r_\text{s} g_\text{s}^2}{\pi^2} \bigg[
		\alpha_0+\alpha_1 {b^\prime}^2+(\alpha_2+\alpha_3 {b^\prime}^2 )\ln{b^\prime}
		\bigg ].
	\end{align}
	With $\eta=\gamma-1-3\ln2$ we have
	\begin{eqnarray}
		\alpha_0&=&6 \ln^3 2-{\frac{1}{3}\ln^4 2 }+\ln^2 2\left({\pi^2\over 3}-14\right) -8\, \text{Li}_4\left(\frac{1}{2}\right)
		\nonumber\\&&+
		{\pi^4\over 12} - {\frac{7}{4}(1+\ln{2})\zeta (3)}+ 4\ln(2)
		\nonumber\\&&+
		\eta \left (2 \ln{2} (\ln{2}-2) +\frac{7 }{4}\zeta (3)\right )
		\nonumber\\
		\alpha_1&=&{11\over 18}+{\pi^2\over 108}+{\ln{2}\over 36}\left ({3\pi^2}-74+8\ln{2}(9 \ln{2}-11)\right )
		\nonumber\\&&
		+\frac{5}{12} \zeta (3)+\eta \left ({\pi^2\over 36}-{1\over 6}+{2\ln{2}\over 3}(\ln{2}-1)\right )
		\nonumber\\
		\alpha_2&=&4 \ln{2} (\ln{2}-2)+\frac{7 \zeta (3)}{2}
		\nonumber\\
		\alpha_3&=&\left ({\pi^2\over 18}-{1\over 3}+{4\ln{2}\over 3}(\ln{2}-1)\right ).\nonumber
\end{eqnarray}

$\Lambda_{2}$, the contribution for $x>1$, is
	\begin{align}
		\label{lamda2H}
		\Lambda_{2}=&\int^\infty_1 v(x) {[S_1^\text{Cy.}(x)]}_{x>1}\; dx
		\nonumber\\
		=&\frac{r_\text{s} g_\text{s}^2}{\pi^2} \bigg [
		\beta_0+\beta_1 {b^\prime}^2+(\beta_2+\beta_3 {b^\prime}^2 )\ln{b^\prime}
		\bigg ]
	\end{align}
	with
	\begin{eqnarray}
		\beta_0&=&-\alpha_0-\frac{\pi^4}{90}
		\nonumber\\
		\beta_1&=&-{\pi^2\over 36}+{\ln{2}\over 36}\left (24\ln2(2-3\ln2)-9{\pi^2}+48\ln{2}+56\right )
		\nonumber\\&&
		-\frac{7}{4} \zeta (3)-\eta \left (\frac{2\ln{2}}{3} (\ln{2}-1)+\frac{\pi^2}{12}\right )
		\nonumber\\
		\beta_2&=&-\alpha_2
		\nonumber\\
		\beta_3&=&-\frac{4\ln{2}}{3} (\ln{2}-1)-\frac{\pi^2}{6}
		\nonumber
	\end{eqnarray}
where ${b^\prime}=b2k_\text{F}$, $\text{Li}_n(z)$ is the polylogarithm function \cite{Abramowitz72}, and $\zeta(s)$ is the Riemann zeta function.
Equations (\ref{lamda1H}) and (\ref{lamda2H}) can be expressed in simpler form by writing the values of the constants as
	\begin{eqnarray}
		\Lambda_{1}&=&-\frac {g_\text{s}^2 r_\text{s}} {\pi ^2}
		\bigg [ 0.766477 + 0.073906\, {b^\prime}^2
		\nonumber\\&&+(-0.583834 + 0.068614\, {b^\prime}^2) \ln({b^\prime})
		\bigg],
		\label{lamda1Hb} \\
		\Lambda_{2}&=&-\frac {g_\text{s}^2 r_\text{s}} {\pi ^2}
		\bigg [0.315846 + 0.798215 \, {b^\prime}^2
		\nonumber\\&&+(0.583834 + 1.361340 \, {b^\prime}^2 ) \ln({b^\prime})
		\bigg ].
		\label{lamda2Hb}
	\end{eqnarray}	
On adding Eqs.\ (\ref{lamda1H}) and (\ref{lamda2H}), major cancellations occur and the analytical form of the correlation energy per particle for a harmonic wire simplifies as
\begin{align}
 \label{exactH}
 \epsilon_\text{c}^\text{Hr.}(b, r_\text{s})=&-\frac{\pi^2}{360}-\frac{b^2}{216 {r_\text{s}}^2} \bigg[6 \left(3+\pi ^2\right) \ln\left(\frac{\pi b}{r_\text{s}}\right)
 \nonumber\\
 &+72 \zeta(3) +3 \gamma\left(3+\pi ^2\right)-2 \pi ^2-42\bigg].
\end{align}

Similarly, the correlation energy per particle for a cylindrical wire is calculated. The details are given in Appendix \ref{Appendix:CorrEnergyCyl}. The final expression for the correlation energy per particle is
\begin{align}
 \label{exactC}
 \epsilon^\text{Cy.}_\text{c}(b, r_\text{s})=&-\frac{\pi^2}{360}- \frac{b^2}{864 {r_\text{s}}^2}\bigg[ 6 \left(3+\pi ^2\right) \ln\left(\frac{\pi  b}{{r_\text{s}}}\right)
 \nonumber\\
 &+72 \zeta (3)+6 \gamma  \left(3+\pi ^2\right)-5 \pi ^2-51\nonumber\\
 &-\pi ^2 \ln (64)-18 \ln(2) \bigg].
\end{align}

\begin{figure}
	\includegraphics[width=8.5cm]{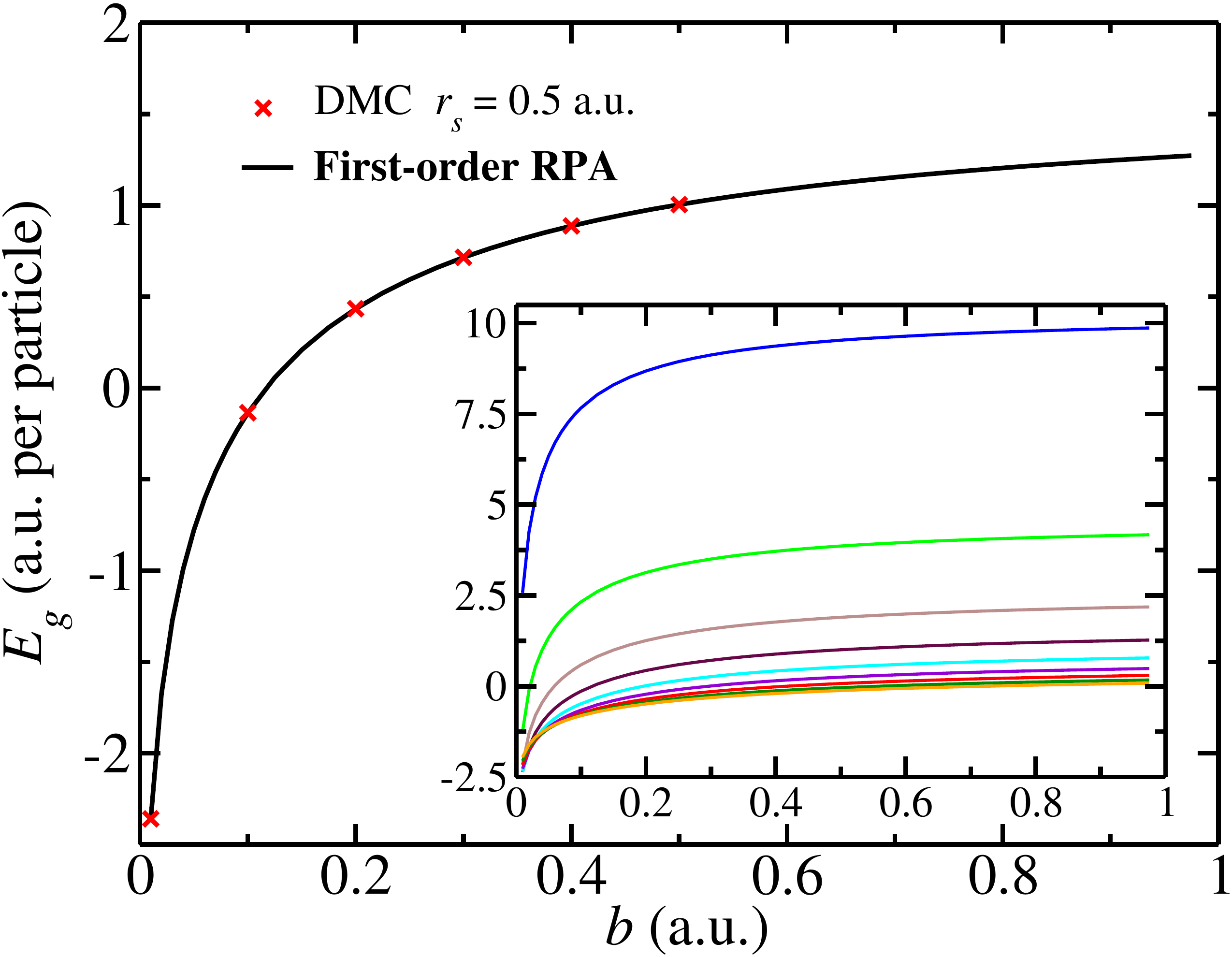}
	\caption{\label{gse_b_dep} Total ground-state energy $E_\text{g}$ as a function of wire width $b$. In the main plot, the  first-order RPA for $r_\text{s}=0.5$ is compared with DMC simulations and in the inset the analytical first-order RPA results are plotted for $r_\text{s}=0.2$--$1$ in steps of 0.1 (top to bottom). The statistical error bars on the DMC results are omitted as they are much smaller than the symbols used.}
\end{figure}

 The first term of Eqs.\ (\ref{exactH}) and (\ref{exactC}) has been found previously using conventional perturbation theory \cite{Loos13,Loos16}, variant RPA \cite{Vinod20}, and QMC \cite{Vinod18c}. Here, we report the next term in the expansion. The $b$ and $r_\text{s}$ dependence enables one to study correlation effects for a finite thickness wire. The expressions for the correlation energy per particle in Eqs.\ (\ref{exactH}) and (\ref{exactC}) are plotted against the wire width $b$ at $r_\text{s}=0.5$, and compared with exact numerical results for both wires in Fig.\ \ref{ce_2nd_order}. The plot shows that the derived expressions are applicable only in the limit $b \ll r_\text{s}<1$.

\begin{figure}
 \includegraphics[width=8.5cm]{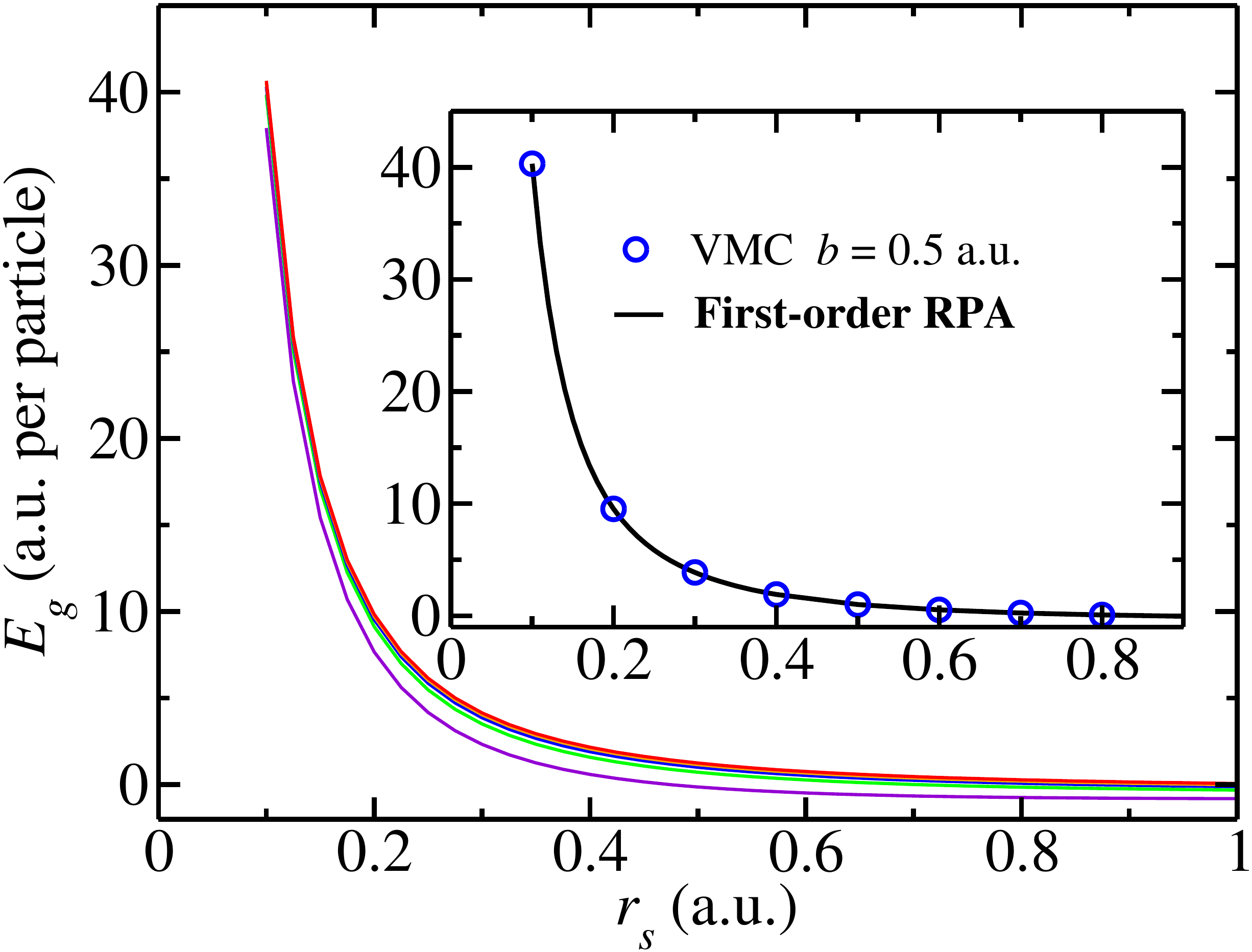}
  \vskip 0.2in
 \includegraphics[width=8.5cm]{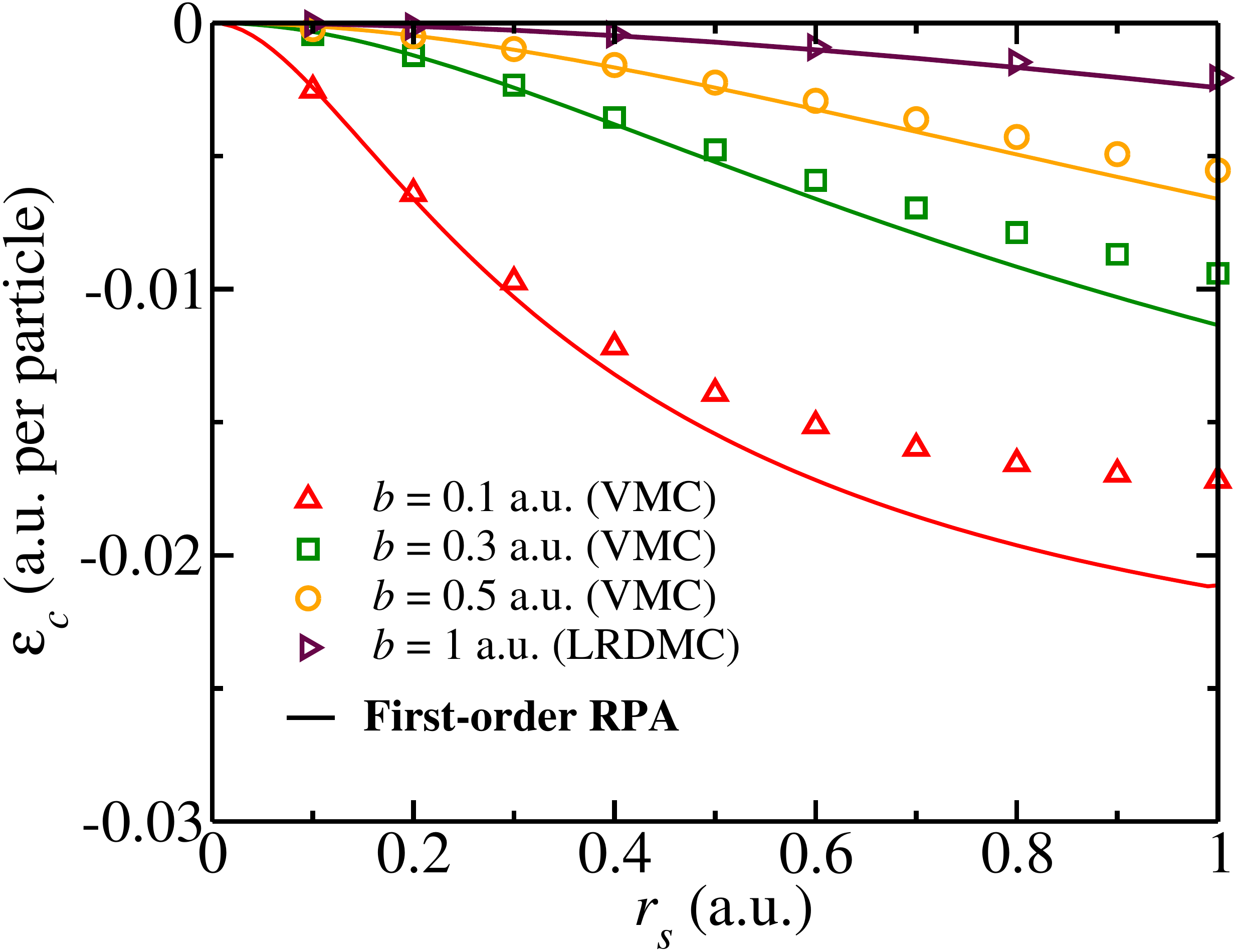}
 \caption{\label{gse_rs_dep} Top panel: total ground-state energy ($E_g$) as a function of density parameter $r_\text{s}$. In the main plot, analytical first-order RPA results are plotted for $b=0.1$, $0.3$, $0.5$, $0.7$, and $0.9$ (bottom to top), and in the inset the first-order RPA results for $b=0.5$ are compared with VMC simulations. Bottom panel: correlation energy per particle for a harmonic wire calculated using first-order RPA (shown as solid lines) compared with the available QMC simulation data. The LRDMC data shown are taken from Ref.\ \cite{Shulenburger08t}}
\end{figure}

Having found all the components of the ground-state energy [Eq.\ (\ref{gse_all})], one can obtain the ground state energy. In Fig.\ \ref{gse_b_dep}, the ground state energy as a function of wire width for various values of $r_\text{s}$ is plotted. It shows that, as the wire width is reduced, the ground state energy decreases and becomes negative for smaller $b$ values, signifying that the 1D HEG is energetically more stable. This stabilization is only for an ideal HEG with a neutralizing background, and neglects the transverse confinement energy. Further, we study the $r_\text{s}$ dependence of the ground state energy and compare it with our recent VMC calculations \cite{Vinod18c} in Fig.\ \ref{gse_rs_dep} (upper), finding excellent agreement. The first-order RPA correlation energy as a function of $r_\text{s}$ is also compared with VMC and lattice-regularized diffusion Monte Carlo (LRDMC) data \cite{Casula06,Shulenburger08,Shulenburger08t} in Fig.\ \ref{gse_rs_dep} (lower). The correlation energy as a function of $r_\text{s}$ for a given fixed thickness $b$ is in good agreement with previous LRDMC simulations. The RPA correlation energy deviates from the QMC correlation energy for smaller values of $b$ and larger values of $r_\text{s}$. The RPA used in the present work is a good approximation for $r_\text{s}<1$. It can be seen from Eq.\ (\ref{harmonic_potential}) that for wires of finite thickness, the effective coupling depends on two length scales, $r_\text{s}$ and $b$. For $r_\text{s}<1$, the coupling increases as $b$ decreases and hence the accuracy of the RPA expansion decreases, leading to the observed deviation from the QMC results.

\section{Conclusions}
\label{Conclusion}
In this paper, we have performed VMC and DMC simulations of the ground-state properties of a finite-width harmonic wire in the high-density regime. The MD data have been used to find the TL parameter $K_\rho$ for several wire widths at high densities. The TL parameter is found to change around 10\% from its value for thin wires. We have also obtained analytical expressions for the wire width-dependent SSF and correlation energy for cylindrical and harmonic models of transverse confinement at high density. Further, we provide numerical results for the wire-width-dependent PCF, SSF, and correlation energy. First-order RPA correlation energies are found to deviate from QMC data for smaller values of $b<r_\text{s}$ due to the fact that the effective electron-electron coupling increases as $b$ decreases in the high density limit.

\begin{acknowledgments}
One of the authors (A.G.)\ is thankful to Ministry of
Human Resource Development (M.H.R.D.)\ for the
financial support to carry out this research work. V.A.\ acknowledge the support in the form of DST-SERB Grant No.\ EEQ/2019/000528. K.N.P.\ acknowledges award of position of honorary scientist by National Academy of Sciences, Prayagraj. The National PARAM Supercomputing Facility (NPSF) at C-DAC, Pune was used for the computational work. K.M.\ acknowledges support from DFG-project
MO 621/28-1.
\end{acknowledgments}

\appendix

\onecolumngrid

\section{}
\label{App:ssf_b_indep}
For reference, we provide here the small-$b$ SSF $S_1(x)$ for harmonic and cylindrical wires, where $x=q/(2k_\text{F})$. For the harmonic wire, it reads for $x<1$ as,
\begin{align}
 \label{SH1a}
 S^\text{Hr.}_1(x)=&\frac{g_\text{s}^2 r_\text{s}}{\pi ^2 x} \big\{-(x+1) \ln ^2(x+1)+2
 (x+1)[\ln (x)+1] \ln (x+1) +(x-1) \ln (1-x)[\ln
 (1-x)-2 (\ln (x)+1)]\big\}
\end{align}
and for $x>1$ as
\begin{align}
 \label{SH1b}
 S^\text{Hr.}_1(x)=&-\frac{g_\text{s}^2 r_\text{s}}{\pi ^2 x} \bigg\{-2 x [\ln (x)+1] \ln
 \left(x^2-1\right)
 +(x-1) \ln ^2(x-1)+\ln ^2(x+1)
 \nonumber\\
 &+ x\left[\ln ^2(x+1)+2 \ln (x) (\ln (x)+2)\right]-4 [\ln
 (x)+1] \coth ^{-1}(x)\bigg\}.
\end{align}
Similarly, for the cylindrical wire,
\begin{align}
&S^\text{Cy.}_1(x)=\frac{g_\text{s}^2 r_\text{s}}{\pi ^2 x}\left \{
\begin{array}{ll}
 \zeta(x) & \text{~if~} x<1
 \cr
 \zeta(x)-2 x \ln x \ln e^2 x & \text{~if~} x>1
\end{array}
\right.
\end{align}
with
\begin{eqnarray}
\zeta(x)&=&(x+1)\ln (x+1)\ln \left( \frac{x^2 e^2}{x+1} \right) +|x-1|\ln |x-1|\ln\left (\frac{x^2 e^2}{|x-1|} \right)
\end{eqnarray}
The first-order SSF in the small $b$ limit comes out to be independent of the width parameter for both the wires.

\section{}

\twocolumngrid

\label{Appendix:CorrEnergyCyl}

	The correlation energy per particle up to second order in $b$ for a cylindrical wire is given by
	\begin{align}
	\epsilon_\text{c}^\text{Cy.}=
	\frac{1}{4r_\text{s}}\bigg\{\Lambda_{1}+\Lambda_{2} \bigg\},
\end{align}
where $\Lambda_{1}$, the contribution for $x<1$, is
\begin{align}
	\label{lamda1C}
	\Lambda_{1}=&\int^1_0 v(x) [S^\text{Cy.}_1(x)]_{x<1}\;dx
	\nonumber\\
	=&\frac{r_\text{s} g_\text{s}^2}{\pi^2} \bigg[
	\alpha_0+\alpha_1 {b^\prime}^2+(\alpha_2+\alpha_3 {b^\prime}^2 )\ln{b^\prime}
	\bigg ].
\end{align}
With $\eta=\gamma-1-3\ln2$ we have
\begin{eqnarray}
	\alpha_0&=&8 \ln^3 2-{\ln^4 2\over 3}+\ln^2 2\left({\pi^2\over 3}-16\right) -8\, \text{Li}_4\left(\frac{1}{2}\right)
	\nonumber\\&&
	+{\pi^4\over 12} +\eta \left (4 \ln{2} (\ln{2}-2) +\frac{7 }{2}\zeta (3)\right )
	\nonumber\\
	\alpha_1&=&{11\over 72}+{\pi^2\over 432}+{\ln{2}\over 9}\left ({\pi^2\over 4}-5+\ln{2}(6 \ln{2}-7)\right )
	\nonumber\\&&
	+\frac{5}{48} \zeta (3)+\eta \left ({\pi^2\over 72}-{1\over 12}+{\ln{2}\over 3}(\ln{2}-1)\right )
	\nonumber\\
	\alpha_2&=&4 \ln{2} (\ln{2}-2)+\frac{7 \zeta (3)}{2}
	\nonumber\\
	\alpha_3&=&\left ({\pi^2\over 72}-{1\over 12}+{\ln{2}\over 3}(\ln{2}-1)\right ). \nonumber
\end{eqnarray}

For $\Lambda_{2}$, the contribution for $x>1$, is
\begin{align}
	\label{lamda2C}
	\Lambda_{2}=&\int^\infty_1 v(x) {[S_1^\text{Cy.}(x)]}_{x>1}\; dx
	\nonumber\\
	=&\frac{r_\text{s} g_\text{s}^2}{\pi^2} \bigg [
	\beta_0+\beta_1 {b^\prime}^2+(\beta_2+\beta_3 {b^\prime}^2 )\ln{b^\prime}
	\bigg ]
\end{align}
with
\begin{eqnarray}
	\beta_0&=&-\alpha_0-\frac{\pi^4}{90}
	\nonumber\\
	\beta_1&=&-{\pi^2\over 144}+{\ln{2}\over 36}\left (14-3{\pi^2}+4 \ln{2}(7-6\ln{2})\right )
	\nonumber\\&&
	-\frac{7}{16} \zeta (3)-\eta \left (\frac{\ln{2}}{3} (\ln{2}-1)+\frac{\pi^2}{24}\right )
	\nonumber\\
	\beta_2&=&-\alpha_2
	\nonumber\\
	\beta_3&=&-\frac{\ln{2}}{3} (\ln{2}-1)-\frac{\pi^2}{24}.\nonumber
\end{eqnarray}

Numerically, Eqs.\ (\ref{lamda1C}) and (\ref{lamda2C}) read
\begin{eqnarray}
	\label{lamda1Cb}
	\Lambda_{1}&=&-\frac {g_\text{s}^2 r_\text{s}} {\pi ^2}
	\bigg [1.00266+0.00296037\, {b^\prime}^2
	\nonumber\\&&-(0.583834-0.0171535\, {b^\prime}^2) \ln({b^\prime})
	\bigg],\\
	\Lambda_{2}&=&-\frac {g_\text{s}^2 r_\text{s}} {\pi ^2}
	\bigg [0.079662-0.108293\, {b^\prime}^2
	\nonumber\\&&+(0.583834+0.340335\, {b^\prime}^2 ) \ln({b^\prime})
	\bigg ].
\end{eqnarray}

\twocolumngrid

\end{document}